\documentclass{aastex63}
\usepackage[parfill]{parskip}
\usepackage{graphicx}
\usepackage{amssymb}
\usepackage{float}
\usepackage{amsfonts}
\usepackage{amsmath}
\usepackage{epstopdf}
\usepackage{amsthm}
\usepackage{tikz}
\usepackage{pgfplots}
\usepackage[english]{babel}
\usepackage{dsfont}
\usepackage[mathscr]{euscript}
\usepackage{bbm}
\usepackage{hyperref}

\let\oldcitet=\citet
\let\oldcitep=\citep 
\let\oldcitealt=\citealt
\renewcommand{\citet}[1]{\textcolor[rgb]{0,0,1}{\oldcitet{#1}}}
\renewcommand{\citep}[1]{\textcolor[rgb]{0,0,1}{\oldcitep{#1}}}
\renewcommand{\citealt}[1]{\textcolor[rgb]{0,0,1}{\oldcitealt{#1}}}
\newcommand{\ms}{\mbox{m s$^{-1}$}}

\received{\today}
\submitjournal{AJ}

\newcommand{\changes}[1]{{\color{black}#1}}

\shorttitle{Stellar Activity F-statistic}
\shortauthors{Holzer et al.}

\graphicspath{{./}{figures/}}

\begin{document}
\title{A Stellar Activity F-statistic for Exoplanet Surveys (SAFE)}

\correspondingauthor{Parker H. Holzer}
\email{parker.holzer@yale.edu}

\author[0000-0001-8936-6276]{Parker H. Holzer}
\affiliation{Department of Statistics \& Data Science, Yale University, 24 Hillhouse Ave., New Haven, CT 06511, USA}

\author[0000-0002-9656-2272]{Jessi Cisewski-Kehe}
\affiliation{Department of Statistics, University of Wisconsin-Madison, 1300 University Ave., Madison, WI 53706,
USA}

\author[0000-0002-3852-3590]{Lily Zhao}
\affiliation{Department of Astronomy, Yale University, 52 Hillhouse Ave., New Haven, CT 06511, USA}

\author[0000-0001-6545-639X]{Eric B. Ford}
\affiliation{Department of Astronomy \& Astrophysics, The Pennsylvania State University, 525 Davey Lab, University Park, PA 16802, USA}
\affiliation{Center for Exoplanets \& Habitable Worlds, The Pennsylvania State University, 525 Davey Laboratory, University Park, PA 16802, USA}
\affiliation{Institute for Computational and Data Sciences, The Pennsylvania State University, 525 Davey Laboratory, University Park, PA 16802, USA}
\affiliation{Penn State Center for Astrostatistics, University Park, PA 16802, USA}
\affiliation{Institute for Advanced Study}

\author[0000-0002-1743-3684]{Christian Gilbertson}
\affiliation{Department of Astronomy \& Astrophysics, The Pennsylvania State University, 525 Davey Lab, University Park, PA 16802, USA}
\affiliation{Center for Exoplanets \& Habitable Worlds, The Pennsylvania State University, 525 Davey Laboratory, University Park, PA 16802, USA}
\affiliation{Institute for Computational and Data Sciences, The Pennsylvania State University, 525 Davey Laboratory, University Park, PA 16802, USA}
\affiliation{Penn State Astrobiology Research Center, University Park, PA 16802, USA}

\author[0000-0003-2221-0861]{Debra A. Fischer}
\affiliation{Department of Astronomy, Yale University, 52 Hillhouse Ave., New Haven, CT 06511, USA}

\begin{abstract}

In the search for planets orbiting distant stars the presence of stellar activity in the atmospheres of observed stars can obscure the radial velocity signal used to detect such planets. Furthermore, this stellar activity contamination is set by the star itself and cannot simply be avoided with better instrumentation. Various stellar activity indicators have been developed that may correlate with this contamination. We introduce a new stellar activity indicator called the Stellar Activity F-statistic for Exoplanet surveys (SAFE) that has higher statistical power (i.e., probability of detecting a true stellar activity signal) than many traditional stellar activity indicators in a simulation study of an active region on a Sun-like star with moderate to high signal-to-noise. Also through simulation, the SAFE is demonstrated to be associated with the projected area on the visible side of the star covered by active regions. We also demonstrate that the SAFE detects statistically significant stellar activity in most of the spectra for HD 22049, a star known to have high stellar variability. Additionally, the SAFE is calculated for recent observations of the \changes{three} low-variability stars \changes{HD 34411, HD 10700,} and HD 3651, the latter of which is known to have a planetary companion. \changes{As expected, t}he SAFE for \changes{these three} only occasionally detects activity. \changes{Furthermore, initial exploration appears to indicate that the SAFE may be useful for disentangling stellar activity signals from planet-induced Doppler shifts.} 

\end{abstract}

\section{Introduction} \label{sxn:intro}
In the exoplanet discovery community focused on extreme precision radial velocity (RV) measurements, better understanding the activity that takes place in the atmospheres of stars is an important goal \citep{fischer16, dumusque17, davis17, jones17, dumusque18, cretignier20}. One reason why this is the case is because stellar activity can produce RV's that mimic, or hide, the signal induced by an orbiting exoplanet \citep{saar97, queloz01, desort07, meunier10}. This is one of the main hinderances to discovering more Earth-like exoplanets around Sun-like stars \citep{hatzes02, lagrange10, isaacson10}. Therefore, to discover more planetary systems similar to the Earth-Sun system, we likely need a better way to detect the presence of, and ideally remove the effects of, stellar activity in stellar spectra.

Many statistics developed by astronomers aim to diagnose stellar activity, and are commonly referred to as ``stellar activity indicators." Some of these are built from the cross-correlation function (CCF) commonly used to derive the RV of the star (e.g., \citealt{baranne96, pepe02}). Several stellar activity indicators, such as the Bisector Inverse Slope (BIS) \citep{queloz01}, the Bisector Slope (BiSlope) \citep{dall06}, Velocity span (V-span) \citep{boisse11}, and Bi-Gaussian (BiGauss) \citep{figueira13}, aim to detect a change in the asymmetry of the CCF, which would represent an asymmetric change in the shape of spectral absorption features on average. Others, such as the Full-Width Half-Maximum of the CCF (FWHM) \citep{queloz01}, instead aim to detect a symmetric change that results in the broadening or narrowing of the CCF. Other developed statistics, not calculated from the CCF, instead look for a change in specific absorption features that are physically known to be sensitive to the stellar magnetic field (e.g., \citealt{queloz09, pont11}), such as emission in the core of the H$\alpha$
line \citep{giguere16}. More recent studies develop indicators that are primarily data-driven \citep{davis17, jones17}.

In this paper, we introduce a new stellar activity indicator developed by modeling the stellar spectrum with linear regression and calculating an F-statistic. This proposed statistic utilizes the idea in \citet{holzer20} that formulated the task of detecting an RV in the spectrum of a star as (weighted) simple linear regression (i.e., linear regression with a single explanatory variable). It accounts for different activity-induced shape changes in different absorption features, allows for straight-forward statistical inference, and is kept general so as to not assume a single predefined shape-change from stellar activity. We also allow this statistic to adapt to the spectral line depths across the spectrum that depend on the chemical abundance, effective temperature, and surface gravitational acceleration of the star observed.

We refer to this new stellar activity indicator as the Stellar Activity F-statistic for Exoplanet surveys (SAFE). It is an F-statistic used in a specially designed least-squares regression model. It is also designed to be robust to changes that are only due to orbiting exoplanets. Particularly, a small Doppler shift of the stellar spectrum would not affect the SAFE. 

In section 2 we introduce the general method of calculating the SAFE, beginning from the perspective of a single absorption feature and then extending to multiple features. The SAFE is then analyzed through simulations of stellar activity in section 3. Additionally, the SAFE is applied to recently collected data from EXPRES \citep{jurgenson16,petersburg2020} in section 4. Finally, a discussion of implications from the SAFE is provided in section 5 and we conclude with section 6.

\section{Method} \label{sxn:methods}
We begin by introducing the linear model used to calculate the SAFE statistic. By design, statistical inference is straight-forward with the SAFE, allowing accurate calculation of a p-value for testing the null hypothesis that there are no effects of stellar activity in the spectrum at the time of observation.

The general idea behind the SAFE is to detect any change over time in the shape of absorption features that is not simply due to a Doppler shift. Such a change is often shared across many absorption features in the spectrum. Examples of such changes would be a change in the asymmetry of the features, an altered line depth, or a broadening or narrowing of the features. Overall, the goal of the SAFE is to test the hypothesis that the only spectrum-wide change in absorption features is a Doppler shift. We note, however, that shape changes of absorption features can also be due to inaccurately correcting for instrumental effects such as the blaze function. Therefore, we assume throughout this paper that corrections for such effects have been done accurately and that the only remaining source of spectrum-wide shape changes is stellar activity.

An essential ingredient for the SAFE is a template spectrum that accurately represents a continuum normalized spectrum for the average state of the star. The need for such a template comes from how the SAFE at time $t$ is calculated from the difference flux at time $t$, defined as 

\begin{equation}
    \mathrm{Diff}_{t}(x) = f_{t}(x) - \tau(x)
    \label{eqn:differenceflux}
\end{equation}

where $x$ is the wavelength of stellar light, $f_{t}$ is the observed normalized flux at time $t$, and $\tau$ is the template flux. If the effective temperature, surface gravitational acceleration, chemical abundances, and other stellar parameters are approximately known, one can use these to generate a synthetic spectrum as a template \citep{sneden12}. However, since these stellar parameters are rarely known to the precision that is likely required for the SAFE, we propose the combined non-parametric smoothing approach of \citet{holzer20}. A second ingredient, which is primarily used in the template estimation, is an RV estimate for each observed spectrum. This estimate could come from the Hermite-Gaussian approach in \citet{holzer20}, a CCF-based approach \citep{mayor95, pepe02, petersburg2020}, or any other method that estimates the RV well\footnote{\changes{We found through a simulation study that the SAFE statistic is robust to RV estimate errors within $5$ \ms\ of the true velocity.}}. Finally, a third ingredient is a wavelength mask giving approximate wavelength intervals corresponding to absorption features. Since we do not make any assumptions about the shape of absorption features, a mask covering as many features as possible is preferred. We use the Absorption Feature Finder (AFF) algorithm described in \citet{holzer20} for this.

This method also requires one to pick a set of basis functions, each representing a particular shape change of absorption features in the spectrum. We propose the generalized Hermite-Gaussian functions defined as
\begin{equation}
\psi_n(x; \mu, \sigma) = \dfrac{1}{\sqrt{\sigma 2^n n!\sqrt{\pi}}} H_n\left(\dfrac{x - \mu}{\sigma}\right) e^{-\dfrac{(x - \mu)^2}{2 \sigma^2}} \label{eqn:genhermgauss_def}  ,
\end{equation}
where $\mu$ is the location parameter, $\sigma$ is the spread parameter, and $H_{n}$ is the $n$'th degree (physicist's) Hermite polynomial which can be expressed as

\begin{eqnarray}
H_{n}(z) = n! \sum\limits_{m=0}^{\lfloor n/2 \rfloor} \dfrac{(-1)^{m}}{m! (n-2m)!} (2z)^{n-2m}
\label{eqn:hermitepoly}
\end{eqnarray} \citep{lanczos38}.

In Equation \eqref{eqn:hermitepoly}, $\lfloor a \rfloor$ represents the floor function that returns the largest integer less than or equal to the real number $a$. By a simple change of variables one can show that the set of generalized Hermite-Gaussian functions forms a complete orthonormal basis of the set of all square-integrable real-valued functions, $L^2(\mathbb{R})$ \citep{johnston14}. Therefore, for a general absorption feature, as long as the difference flux due to either a Doppler shift or stellar activity is such an $L^2(\mathbb{R})$ function, it can be decomposed into the generalized Hermite-Gaussian function basis. We do acknowledge, however, that there are many other choices that satisfy these properties. We prefer the Hermite-Gaussian basis because it is composed of differentiable functions that are either even or odd (with respect to $\mu$) and are localized like spectral lines. Another motivation for this basis is that many absorption features are close to being Gaussian-shaped, and the Hermite-Gaussian basis contains both the Gaussian and its derivative.

\subsection{A Single Absorption Feature} \label{subsxn:single_feature}

A large portion of the spectrum is used to calculate the SAFE; however, for simplicity we begin by considering the scenario of a single absorption feature indexed by $i$.

\citet{holzer20} show mathematically that if an absorption feature is Gaussian-shaped, the difference flux due to a Doppler shift can be well modeled by only using $\psi_{1}$. However, since absorption features are often strongly blended, and the wings of large features are more Lorenzian than Gaussian, we wish to relax the assumption of Gaussianity. To do so we instead assume that the difference flux can be written as a linear combination of all Hermite-Gaussian functions up to degree five\footnote{The degree $5$ limit was selected to make the SAFE flexible enough to detect a variety of possible shape changes of absorption features, but not so flexible that it affects the statistical power of the test due to too many additional variables in the model. We found that up to degree $5$ provides a good balance of these motivations.}. This assumption helps to relax the assumption made in \citet{holzer20} that absorption features are Gaussian (i.e., that the difference flux can be written as a constant multiple of just $\psi_{1}$). Written in equation form, we model the difference flux due to a Doppler shift of absorption feature $i$, $\mathrm{Doppdiff}_{i}$, as
\begin{equation}
\mathrm{Doppdiff}_{i}(x) = \sum\limits_{d=0}^{5} \alpha_{i,d} \psi_{d}\left( x ; \hat{\mu}_{i}, \hat{\sigma}_{i} \right) \label{eqn:singlefeature_doppmod}
\end{equation}
 where $\hat{\mu}_{i}$ and $\hat{\sigma}_{i}$ are the location and spread estimated from fitting a Gaussian to feature $i$.
 
 In practice, to get estimates $\hat{\mu}_{i}$ and $\hat{\sigma}_{i}$, we follow the approach in \citet{holzer20} with a template spectrum for the star. The AFF algorithm is used to identify wavelength windows corresponding to absorption features in the template. Using nonlinear optimization, a sum of three Gaussian density functions are used to fit each absorption feature and its neighboring features. As a further step to correct for inaccurate continuum normalization, the continuum value from which the fitted Gaussian is subtracted, $\mathrm{cont}_{i}$, is also a free parameter (within the bounds of $0.9$ and $1.1$). Features where the optimization does not converge are eliminated. Occasionally, the optimization converges to a poor fit. Therefore, any feature is eliminated if $\hat{\sigma}_{i}$ is more than $1.5$ multiples of the interquartile range above the $75$'th percentile of all the fitted $\hat{\sigma}_{i}$, $\hat{\sigma}_{i}$ is less than $0.001$ \AA, $\hat{\sigma}_{i}$ is greater than one sixth of the feature wavelength window width, or the amplitude of the fitted Gaussian is positive. Finally, any feature where $\widehat{\mathrm{cont}}_{i} < 0.98$ or $\widehat{\mathrm{cont}}_{i} > 1.02$ is also eliminated. On average, we find that approximately $45$ percent of the features identified by the AFF algorithm remain after these cuts.
 
 If we knew how to write $\mathrm{Doppdiff}(x)$ in closed form, we could solve for the coefficients $\alpha_{i,d}$ exactly. However, since this is not the case, we use least-squares to project the discretized realization of $\mathrm{Doppdiff}_{i}(x)$ onto the space spanned by $\lbrace \psi_{d}\left( x ; \hat{\mu}_{i}, \hat{\sigma}_{i} \right) \rbrace_{d=0}^{5}$. This gives $\hat{\alpha}_{i,0}$, $\hat{\alpha}_{i,1}$, $\hat{\alpha}_{i,2}$, ..., $\hat{\alpha}_{i,5}$ as estimates of the projection coefficients, and $\widehat{\mathrm{Doppdiff}}_{i}$ as the projected difference flux due to a Doppler shift.

In practice, we train $\widehat{\mathrm{Doppdiff}}_{i}$ by taking the template spectrum for absorption feature $i$, simulate a $10$ \ms\ Doppler shift, interpolate back to the original wavelengths, and use least-squares to project the resulting difference flux onto the generalized Hermite-Gaussian functions of degree $5$ or lower. $\widehat{\mathrm{Doppdiff}}_{i}(x)$ is similar to an estimate of the $i$'th absorption feature's derivative. Therefore, for any small RV the difference flux due to a Doppler shift is approximately a constant multiple of $\widehat{\mathrm{Doppdiff}}_{i}(x)$. 

If stellar activity is present, then a Doppler shift is not the only source of the difference flux. Therefore, we would expect to have signal remaining in the residual difference spectrum, the difference flux remaining after removing a constant multiple of $\widehat{\mathrm{Doppdiff}}_{i}$. To test for this we model $\mathrm{Diff}_{t}$ defined in Equation \eqref{eqn:differenceflux} as
\begin{equation}
\mathrm{Diff}_{t} (x) = \beta_{1,i,t} \widehat{\mathrm{Doppdiff}}_{i}(x)\ +\ \sum\limits_{\substack{d=0\\ d \neq 1}}^{5} \beta_{d,i,t} \psi_{d}\left( x ; \hat{\mu}_{i}, \hat{\sigma}_{i} \right)\ +\ \varepsilon_{t} \label{eqn:singlefeature_mod}
\end{equation}
%
where the random error $\varepsilon_{t}$ has mean $0$. This reduces the problem to a linear model of the difference flux as a linear combination of a small Doppler shift and the Hermite-Gaussian functions. One may notice that $\psi_{1}\left( x ; \hat{\mu}_{i}, \hat{\sigma}_{i} \right)$ is not included in the model of Equation \eqref{eqn:singlefeature_mod}. Were $\psi_{1}\left( x ; \hat{\mu}_{i}, \hat{\sigma}_{i} \right)$ to be included, the coefficients would be unidentifiable as $\widehat{\mathrm{Doppdiff}}_{i}(x)$ could then be written as a linear combination of the other terms. Furthermore, since most absorption features are approximately Gaussian and therefore $\widehat{\mathrm{Doppdiff}}_{i}(x)$ is almost the same as a first-degree Hermite-Gaussian function, including $\psi_{1}\left( x ; \hat{\mu}_{i}, \hat{\sigma}_{i} \right)$ would potentially lead to collinearity issues.

We furthermore assume that $\varepsilon_{t}$ is multivariate normal with a diagonal covariance matrix whose entries are well approximated by the squared uncertainties of the normalized flux. If these uncertainties are not provided we use the normalized template flux divided by the continuum used for normalization as the diagonal entries. This assumes that the raw flux, $F_{i,t}$, follows a Poisson$(\lambda_{i,t})$ distribution. Under this assumption, for large values of $\lambda_{i,t}$, $F_{i,t}$ approximately follows the Normal$(\lambda_{i,t}, \lambda_{i,t})$ distribution. Therefore, the normalized flux $F_{i,t}/c_{i,t}$ approximately follows the Normal$(\lambda_{i,t}/c_{i,t}, \lambda_{i,t}/c_{i,t}^{2})$ distribution. Since $\lambda_{i,t}/c_{i,t}$ is well-approximated by the template flux $\tau_{i}$ at each time $t$, the difference flux at pixel $i$ approximately follows the Normal$(0, \tau_{i}/c_{i,t})$ distribution. We acknowledge that a diagonal covariance matrix does not account for correlations due to the point spread function of the instrument, a potential improvement to be made in future work.

Now that we have the linear model in Equation \eqref{eqn:singlefeature_mod}, we can use it to perform statistical inference about the null hypothesis that no stellar variability effects are present at time $t$. Testing for the absence of stellar variability reduces to testing whether $\beta_{0,i,t} = \beta_{2,i,t} = \beta_{3,i,t} = \beta_{4,i,t} = \beta_{5,i,t} = 0$ in Equation \eqref{eqn:singlefeature_mod}. This can be done at each time $t$ by comparing the residual sum of squares (RSS) when all five of these coefficients are used in the fit to the RSS when they are set to $0$. These two quantities can be combined together to give an F-statistic that we can use to perform an F-test of the hypothesis that the only nonzero coefficient is $\beta_{1,i,t}$.

\subsection{Multiple Absorption Features} \label{subsxn:multiple_features}

Since the signals of stellar activity and low-mass exoplanets are so small, we need to extend the model to include nearly all absorption features in the spectrum. Using $\mathrm{Doppdiff}(x)$, the difference flux due to a Doppler shift for multiple absorption features in the spectrum, our Doppler model from Equation \eqref{eqn:singlefeature_doppmod} becomes
\begin{equation}
\mathrm{Doppdiff}(x) = \sum\limits_{i=1}^{m} \sum\limits_{d=0}^{5} \alpha_{i,d} \psi_{d}\left( x ; \hat{\mu}_{i}, \hat{\sigma}_{i} \right)\mathbbm{1} \lbrace \hat{l}_{i} \leq x \leq \hat{u}_{i} \rbrace \label{eqn:full_doppmod}
\end{equation}
where $\hat{l}_{i}$ and $\hat{u}_{i}$ are the wavelength bounds for feature $i$ (with no overlap between features), and $\mathbbm{1}\lbrace A \rbrace$ represents the indicator function that takes the value $1$ if the statement $A$ is true and $0$ otherwise. In Equation \eqref{eqn:full_doppmod}, the outer sum is over all $m$ absorption features and the inner sum is over the Hermite-Gaussian degrees up to $5$. To simplify the process of estimating the values for each coefficient $\alpha_{i,d}$, the indicator function is included so that we can perform a least-squares fit on each absorption feature separately as described in Section \ref{subsxn:single_feature}. It also serves to exclude the continuum which has little relevant Doppler information. This, however, requires the assumption that the Doppler shift after applying the barycentric correction is not too large. Without an assumption of spectral line shapes, we are unable to mathematically quantify how small of a RV is small enough, but simulation suggests that less than $100$ \ms\ is acceptable.

Once we obtain $\hat{\alpha}_{i,d}$ as least-squares estimates of the coefficients in Equation \eqref{eqn:full_doppmod}, we obtain $\widehat{\mathrm{Doppdiff}}(x)$. Since this is trained using a Doppler shift that is small ($10$ \ms), $\widehat{\mathrm{Doppdiff}}(x)$ is again an approximation of the template spectrum's derivative. So to expand our model to include all absorption features we simply need to use this term and sum over absorption features for the Hermite-Gaussian terms:
\begin{equation}
\mathrm{Diff}_{t} (x) = \beta_{1,t} \widehat{\mathrm{Doppdiff}}(x)\ +\ \sum\limits_{i=1}^{m} \sum\limits_{\substack{d=0\\ d \neq 1}}^{5} \beta_{d,i,t} \psi_{d}\left( x ; \hat{\mu}_{i}, \hat{\sigma}_{i} \right)\mathbbm{1} \lbrace \hat{l}_{i} \leq x_{t} \leq \hat{u}_{i} \rbrace\ +\ \varepsilon_{t} \label{eqn:full_mod1} .
\end{equation}
Once again, this is a linear model and we can test whether $\forall_{d \leq 5,\ d\neq 1}\ \forall_{i \leq m}\ \beta_{d,i,t} = 0$. However, this is a very high-dimensional test. Since this is likely to cause the power of the test to be too small, we wish to revise the model in Equation \eqref{eqn:full_mod1} so that the dimension of the test is reduced. 

We reduce the dimension by noting that the amplitude of the difference flux will differ across absorption features, but that the relative amplitudes are in many cases consistent. For example, \citet{holzer20} show that when the difference flux is caused by a small Doppler shift, the relative amplitudes of $\psi_{1}(x; \mu, \sigma)$ for Gaussian absorption features are given by an expression involving the line depth and width parameters, both of which can be estimated from the template. Here, we wish to fine-tune the Hermite-Gaussian terms in Equation \eqref{eqn:full_mod1} to effects of stellar activity rather than a Doppler shift. Since we do not assume absorption features are Gaussian, and because there is no existing parametric model for the difference flux from stellar activity, we need to estimate the relative amplitudes statistically rather than mathematically.

For this we use a set of 25 Spot Oscillation and Planet 2.0 (SOAP) spectra \citep{dumusque14}, designed to approximately represent the stellar activity effects of an active region rotating around the Sun. These SOAP spectra are set to have a $1\%$ equatorial active region that starts from the back of the star and rotates around exactly once. They cover the wavelength range from $4470$ to $6662$ \AA\ and have the full resolution of the National Solar Observatory (NSO) spectrum \citep{rimmele98}. Using the first of these spectra as the template, which has no activity effects present since the active region is on the back of the star, we use the AFF algorithm in \citet{holzer20} to identify wavelength bounds of absorption features. We then estimate the Gaussian fit parameters for each feature and project the difference flux due to a $10$ \ms\ Doppler shift onto the space spanned by $\psi_{0}, \psi_{1}, ... , \psi_{5}$ according to Equation \eqref{eqn:full_doppmod} to get the estimated $\widehat{\mathrm{Doppdiff}}(x)$ for the represented star. For each of the 25 SOAP spectra we then subtract out the multiple of $\widehat{\mathrm{Doppdiff}}(x)$ that minimizes the sum of squared residuals. The residual difference flux of a given feature is then projected onto the space spanned by $\psi_{0}, \psi_{2}, ..., \psi_{5}$ as in Equation \eqref{eqn:full_mod1} and the fitted coefficients for each of the $m$ absorption features and each of the $25$ SOAP epochs are stored.

This in turn gives $5$ different $25 \times m$ arrays of SOAP coefficients, each corresponding to a particular Hermite-Gaussian degree. For each of these $5$ arrays, the corresponding $25 \times m$ array can be viewed as $25$ points in $m$ dimensional space. We then use robust principal components analysis (rPCA) \citep{candes11} which decomposes each $25\times m$ matrix into a sum of a low-rank matrix, $L$, and a sparse matrix, $S$, allowing us to avoid being influenced by strong outliers.\footnote{If no strong outliers are present, this gives essentially the same results as the classical principal components analysis.} Taking the singular-value decomposition of $L$ gives us the first principal vector, the loadings $\hat{\gamma}_{d,i}$ of which estimate the relative amplitudes due to an active region. 

\begin{figure}[ht!]
    \centering
    \includegraphics[scale=0.85]{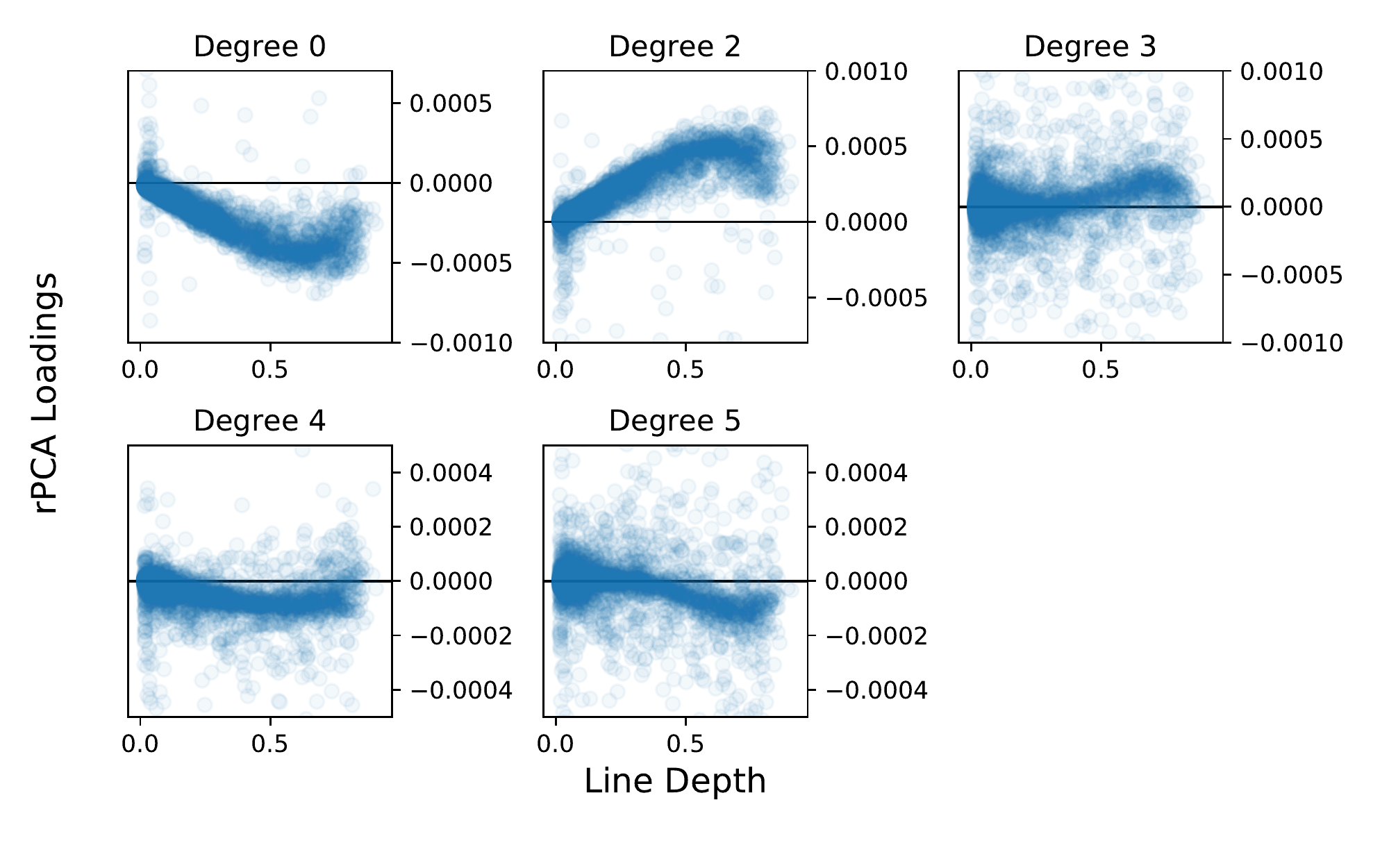}
    \caption{The relative amplitudes of absorption features for each Hermite-Gaussian degree. The amplitudes were obtained by applying rPCA to the Hermite-Gaussian coefficients fitted to the difference flux of SOAP spectra with a $1\%$ equatorial active region. Each panel represents a different Hermite-Gaussian degree. The value of the relative amplitudes are given on the right of each panel, with the horizontal axis representing the line depth of the absorption feature with respect to the normalized flux $1.0$.}
    \label{fig:relative_amps}
\end{figure}

Our approach towards obtaining these relative amplitude estimates is similar to the methods of \cite{davis17}. However, instead of applying principal components analysis to the pixels, our approach applies it to the fitted Hermite-Gaussian coefficients. This allows us to account for the smooth functional relationship of neighboring pixels that belong to the same absorption feature. Furthermore, we only use the first principal vector for each Hermite-Gaussian degree. The set of relative amplitudes for each absorption feature from our approach is similar to the weighted mask used in the CCF method. However, there is more than just a single mask produced in our approach, each is obtained from dimensional reduction of SOAP simulations, and weights are replaced with relative amplitudes. Figure \ref{fig:relative_amps} gives a visualization of the relative amplitudes versus line depth.

Figure \ref{fig:relative_amps} illustrates that the shape change due to stellar activity is related to the depth of the absorption line \citep{cretignier20}. For degrees $0$ and $2$ the amplitude of the shape change appears to be roughly proportional with line depth, until reaching a depth of about $0.6$. At this point the amplitude begins to decrease. Higher degrees also seem to show a relationship with depth, though not as clearly.

Using the estimates of these amplitudes, $\hat{\gamma}_{d,i}$, together with $\hat{\mu}_{i}$ and $\hat{\sigma}_{i}$ that are estimated through preprocessing, we define

\begin{equation}
\hat{\Psi}_{d}(x) := \sum\limits_{i=0}^{m} \hat{\gamma}_{d,i} \psi_{d}\left( x ; \hat{\mu}_{i}, \hat{\sigma}_{i} \right)\mathbbm{1} \lbrace \hat{l}_{i} \leq x \leq \hat{u}_{i} \rbrace\ \label{eqn:dim_reduced_var}.
\end{equation}

Our final dimensionally-reduced model then becomes

\begin{equation}
\mathrm{Diff}_{t} (x) = \beta_{1,t} \widehat{\mathrm{Doppdiff}}(x)\ +\  \sum\limits_{\substack{d=0\\ d \neq 1}}^{5} \beta_{d,t} \hat{\Psi}_{d}(x) +\ \varepsilon \label{eqn:full_mod2} .
\end{equation}

which is again a linear model. To account for the heteroskedasticity of the noise, we use weighted least-squares with the weights set to be the reciprocal of the squared uncertainty in normalized flux, or (if this uncertainty is not provided) the normalization continuum divided by the template flux. 

Equation \eqref{eqn:full_mod2} is again a linear model that can be used for statistical inference about whether or not the difference flux at time $t$ is (at least partially) due to stellar variability. The general idea behind the test statistic built for this purpose is to consider two different models: one where only the first term of Equation \eqref{eqn:full_mod2}, $\widehat{\mathrm{Doppdiff}}$, is included in the model and a second where all the terms are included. If no stellar variability is present at time $t$, then the RSS of the first of these two models will not be significantly greater than the RSS of the second. Specifically, we define our test statistic as 

\begin{equation}
\mathrm{SAFE_{t}} := \dfrac{\mathrm{RSS}_{\mathrm{dopp},t} - \mathrm{RSS}_{\mathrm{full},t}}{\mathrm{RSS}_{\mathrm{full},t}} \dfrac{n - 6}{5} \label{eqn:SAFE_def}
\end{equation}

where $\mathrm{RSS}_{\mathrm{dopp}}$ is the residual sum of squares when using only $\widehat{\mathrm{Doppdiff}}(x)$ in the regression, $\mathrm{RSS}_{\mathrm{full}}$ is the residual sum of squares with all the terms in Equation \eqref{eqn:full_mod2}, and $n$ is the number of data points included in the regression.\footnote{Note that $\left(\mathrm{RSS}_{\mathrm{dopp},t} - \mathrm{RSS}_{\mathrm{full},t}\right)/5$ follows the $\chi^{2}(5)$ distribution and is independent of $\left(\mathrm{RSS}_{\mathrm{full},t}\right)/(n-6)$, which follows the $\chi^{2}(n-6)$ distribution. Therefore, $\mathrm{SAFE_{t}}$ is a ratio of independent chi-squared random variables. This, therefore, can be easily generalized for situations where the covariance matrix of $\varepsilon$ in Equation \eqref{eqn:full_mod2} is not diagonal due to effects like the points spread function.} Under the null hypothesis, SAFE$_t$ would follow the F-distribution with $(5, n-6)$ degrees of freedom, which allows us to calculate a p-value.

\subsection{Multiple Testing Correction}\label{subsxn:multiple_testing}

Since each star is observed multiple times, each of which gives a different value of the SAFE, we need to account for multiple testing when analyzing the p-values given.\footnote{Note that this is not the same as testing the single hypothesis that none of the spectra have stellar activity signals.} For a star observed $N$ different times, we use the Bonferroni correction \citep{bonferroni36} to control the family-wise error rate. For a given significance level, $\alpha$, this entails comparing all $N$ p-values to $\alpha/N$ instead of $\alpha$ when deciding whether or not to reject the null hypothesis of no stellar activity being present. This correction ensures that the probability of falsely concluding the presence of statistically-significant stellar activity in any of the $N$ observations is less than $\alpha$.

For clarification, we note that in this context the p-value is the probability, when no stellar activity is present in the star, of obtaining a spectrum from the target star that gives a larger value of the SAFE than what was actually observed. We also note that when the p-value is found to not be below $\alpha/N$, we do not conclude that there is no stellar activity present. Instead, we simply conclude that, whatever stellar activity might be present, we do not have sufficient evidence to detect it.

\section{Simulation Studies} \label{sxn:simulations}
To test the proposed method, we begin by using a set of spectra simulated from SOAP. Our simulated sample consists of a time-series of $25$ observations of the solar spectrum. The sample includes a $1\%$ equatorial active region that begins at the back of the star and, at each successive observation, rotates around the star by a fixed angle. For the first seven, and the last six, epochs the active region is behind the star and, therefore, the spectrum is the same as the quiet solar spectrum.

\subsection{SAFE Null Distribution} \label{subsxn:safe_null}

Since we have full knowledge of when the active region is present or not in SOAP, we use this knowledge to check the null-distribution of SAFE and to estimate the power of the SAFE test\footnote{The null distribution of the SAFE is the probability distribution it follows when no stellar activity is present (i.e., the null hypothesis is indeed correct). Since the SAFE is an F-statistic, this null distribution is an F-distribution. The power is the probability of concluding that the SAFE is statistically significant when there is indeed stellar activity present.}. To simulate data for a true null, we use one of the SOAP spectra where the active region is behind the star. We begin by creating $13000$ simulations of adding independent Poisson noise to the spectrum that gives a Signal-to-Noise Ratio per pixel (SNR) of $150$. We use the quiet solar spectrum from SOAP as a template and follow the procedure described in Section \ref{sxn:methods}. The AFF algorithm finds a total of $4,114$ absorption features that remain after filtering out those for which convergence of Gaussian fit parameters was not reached or the spread was outside a reasonable range. After estimating the $\mathrm{Doppdiff}$ variable according to Equation \eqref{eqn:full_doppmod}, we use the SOAP amplitudes obtained above through rPCA to define all other covariates of the model in Equation \eqref{eqn:full_mod2}. The estimated Gaussian fit parameters are obtained through the process described near the beginning of Section \ref{subsxn:single_feature}, using the quiet SOAP spectrum as the template.

\begin{figure}[ht!]
\centering
\includegraphics[scale=0.78]{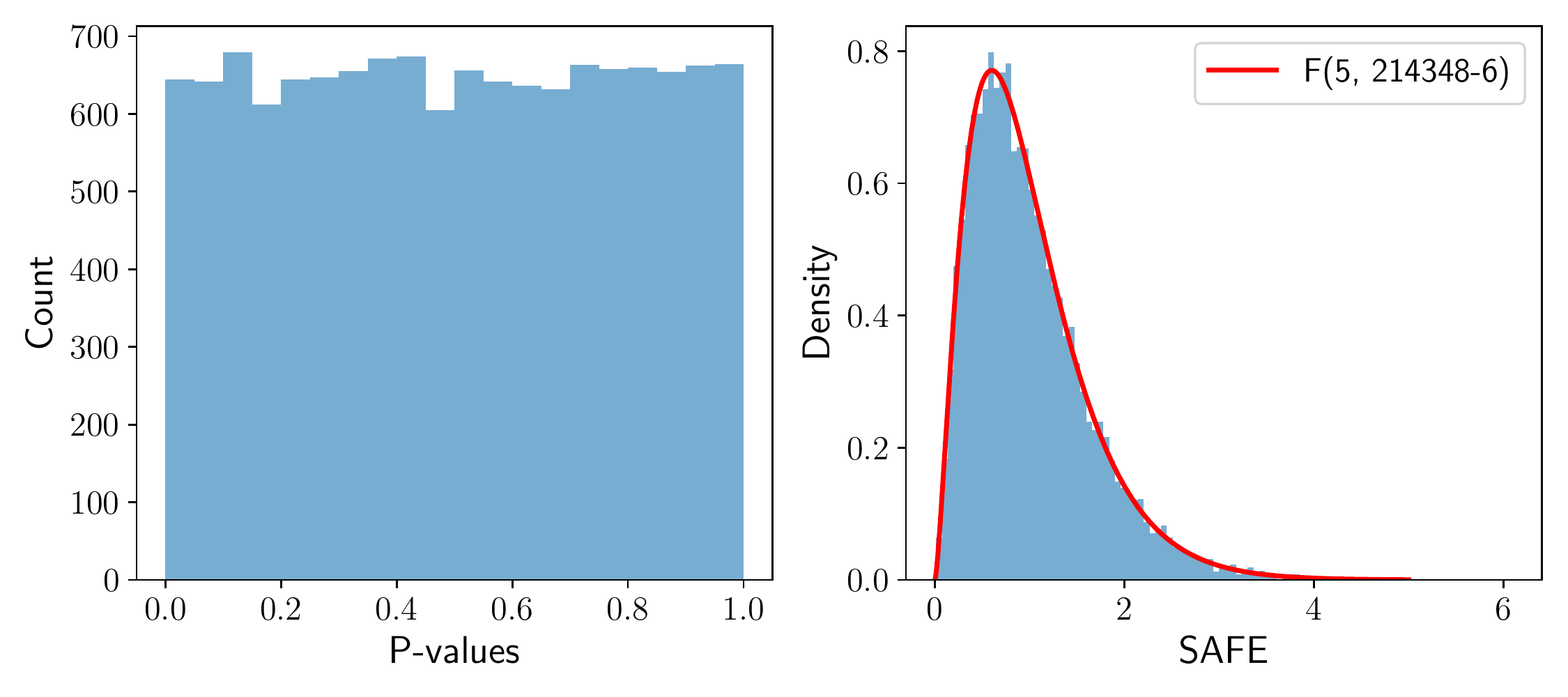}
\caption{The left panel shows the distribution of p-values calculated with SAFE from $13000$ independent simulations of a quiet SOAP spectrum with SNR $150$. The right panel shows the distribution of the SAFE. Under the null hypothesis of no stellar activity, which is true here, the SAFE follows the F$(5, n-6)$ indicated by the solid red curve and the calculated p-values should be Uniform$(0,1)$, which is supported by the results.}
\label{fig:pvalue_null}
\end{figure}

For each simulation we calculate SAFE and it's associated p-value with the proper F-distribution. Since each set of simulated spectra has different noise realizations, each will produce a different p-value. If the distribution of the SAFE under no stellar activity is indeed the F-distribution used to calculate each p-value, then the p-values will follow the Uniform($0,1$) distribution. The distribution of the p-values given by this simulation study is illustrated in the histogram of Figure \ref{fig:pvalue_null}. Since this distribution is approximately uniform, this supports the F$(5, n-6)$ as the null distribution of SAFE (i.e., the distribution that describes the behavior of the SAFE when white noise and a small Doppler shift are the only sources of variation in the spectrum). If the true distribution of the SAFE, under no stellar activity, had a fatter right tail or were shifted upwards more than the F$(5, n-6)$ distribution, the p-values in Figure \ref{fig:pvalue_null} would be more concentrated close to $0.0$ than $1.0$. Similarly, if the right tail of the true null distribution decreased faster than the F$(5, n-6)$ tail, the p-values in Figure \ref{fig:pvalue_null} would be more concentrated near $1.0$.

\begin{figure}[ht!]
\centering
\includegraphics[scale=0.78]{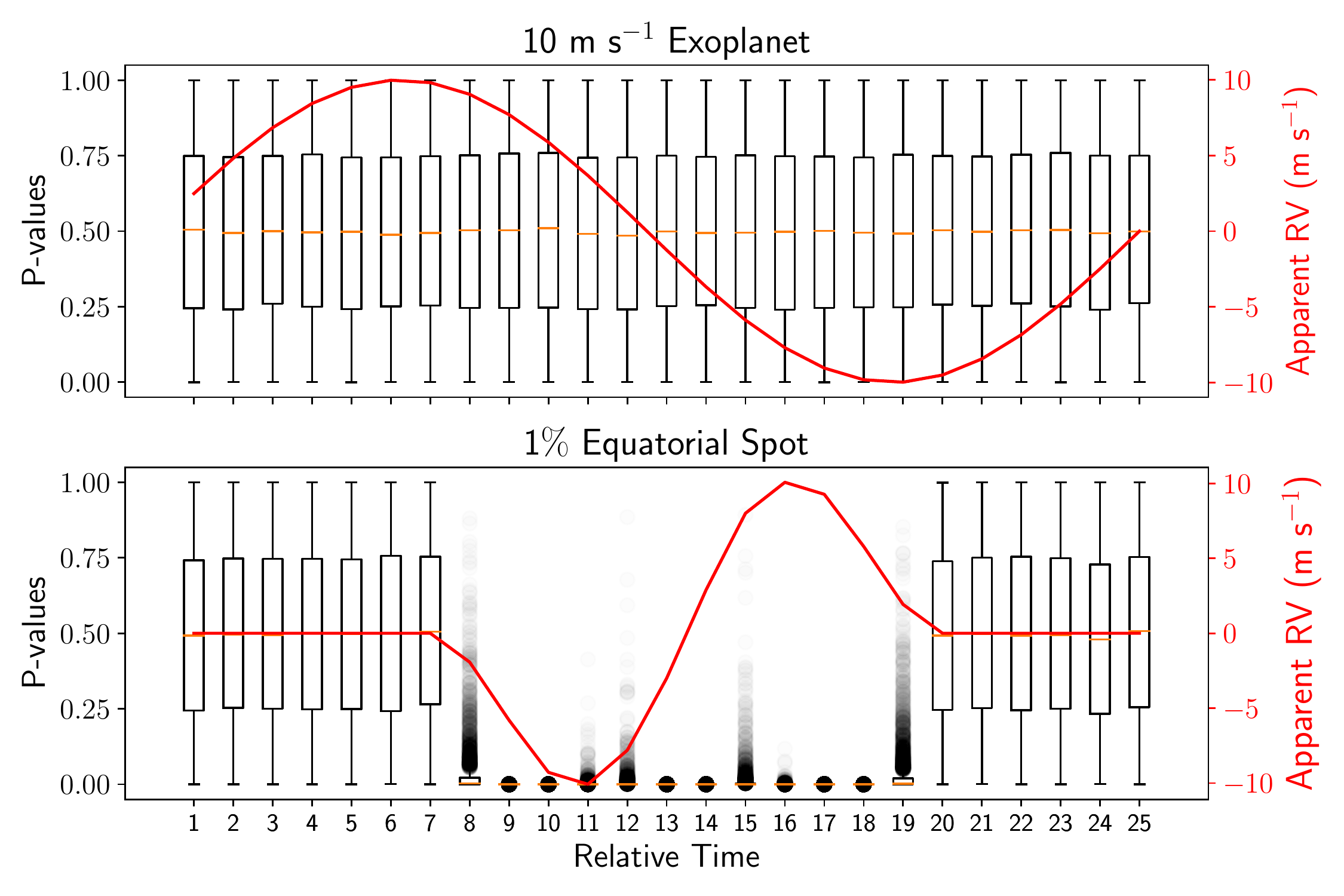}
\caption{Results of doing $5,000$ simulations of adding SNR $150$ noise to the spectrum at each time instance. The top panel is for the case of the only effect being due to a $10$ \ms\ semi-amplitude Doppler shift from an exoplanet going through one complete circular orbit. The bottom panel is for the case of a $1\%$ equatorial active region rotating around the star. At each relative time instance for both cases, the distribution of p-values is shown according to the left vertical axis. The detected radial velocity for each case is indicated with a smooth curve according to the right vertical axis as a way to indicate the phase of the orbit/rotation. The black circles indicate outliers, defined as points that are above the $75$'th percentile by more than $1.5$ multiples of the interquartile range of the p-values at the specified time. }
\label{fig:exo_vs_spot}
\end{figure}

To check if the SAFE still follows the theoretical F-distribution in the presence of a Doppler shift, we follow the same procedure with a time-dependent RV  $v_{t} = 10 \text{sin}(2\pi t/25)$ for $t = 1,2,...,25$, which simulates the effect of an exoplanet completing exactly one circular orbit around the Sun. In a separate analysis we use the $25$ SOAP spectra that represent a $1\%$ equatorial active region. For each of the $25$ phases in both of these data sets, we run $5000$ independent simulations of adding SNR $150$ Poisson noise to the spectrum and calculate SAFE and its associated p-value. Figure \ref{fig:exo_vs_spot} illustrates the distribution of p-values at each phase for both the true Doppler shift and the $1\%$ equatorial active region.

For the simulated data that represents a true Doppler shift, the distribution of p-values should be approximately uniform at every time instance as the null hypothesis is true regardless of the RV. The top panel in Figure \ref{fig:exo_vs_spot} illustrates that this is indeed true. The data representing the equatorial active region begins and ends with the null hypothesis being true as the active region is behind the star. But once the active region comes around the star the p-values should become more concentrated near zero. This scenario is also represented in Figure \ref{fig:exo_vs_spot}.

\subsection{SAFE Power} \label{subsxn:safe_power}

\begin{figure}[ht!]
\centering
\includegraphics[scale=0.75]{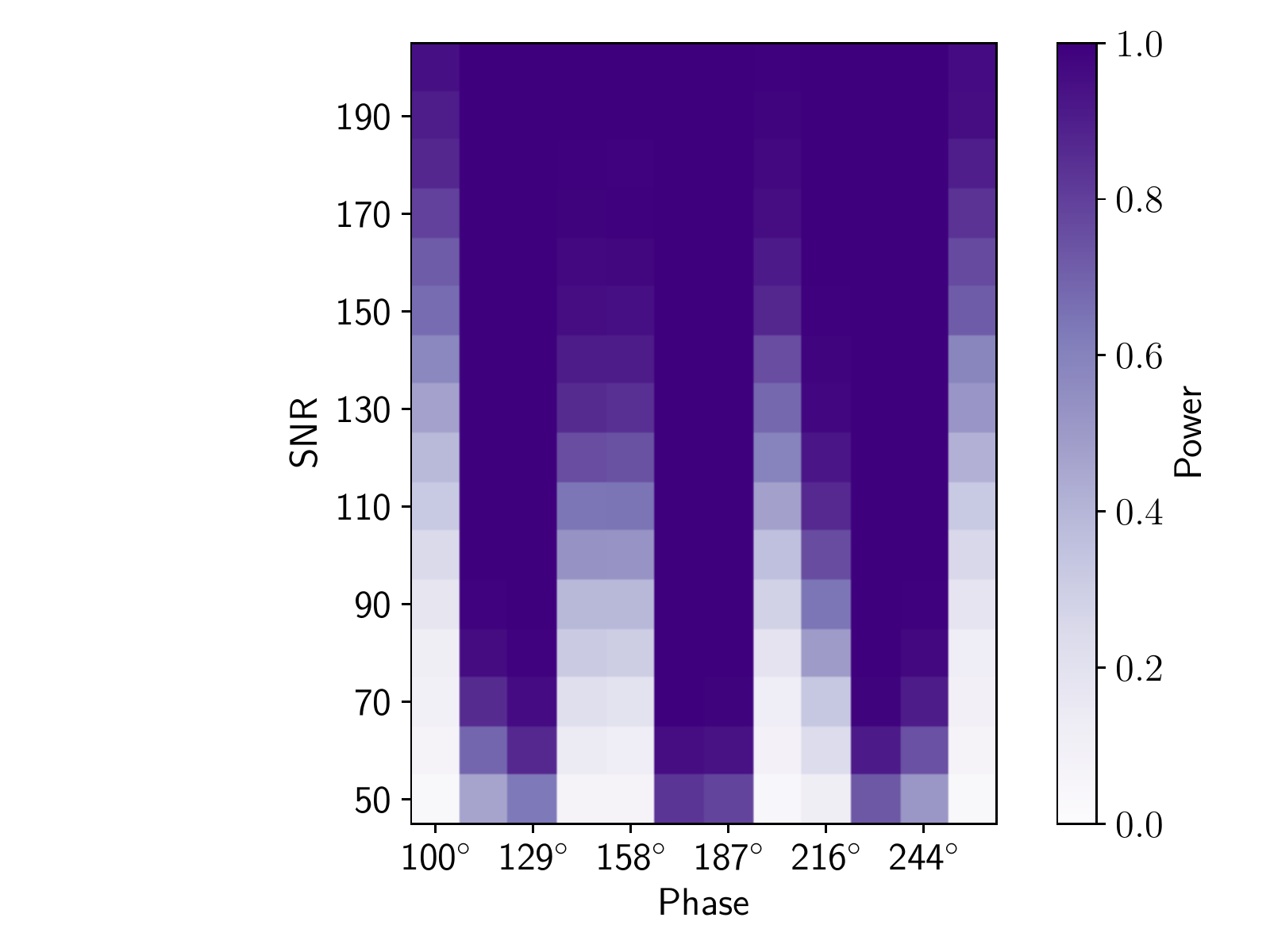}
\caption{The power of the SAFE test, estimated by $1000$ Poisson noise simulations, is plotted as a function of SNR and active region phase. The phases where the active region is behind the star are not included. The power is indicated by the colorbar on the right.}
\label{fig:power_vs_snrphase}
\end{figure}

We investigate the power of the SAFE test as a function of the active region's phase, and how it changes for different levels of noise. For each phase beginning when the active region comes around the star's edge to where it again goes behind, and for each SNR from $50$ up to $200$ in steps of $10$, we add independent Poisson noise to the spectrum and calculate the SAFE and its corresponding p-value from the F-distribution $1000$ times. We then estimate the power from the proportion of p-values that are below the $0.01$ significance level.  The results are shown in Figure \ref{fig:power_vs_snrphase}.

\begin{figure}[ht!]
\centering
\includegraphics[scale=0.58]{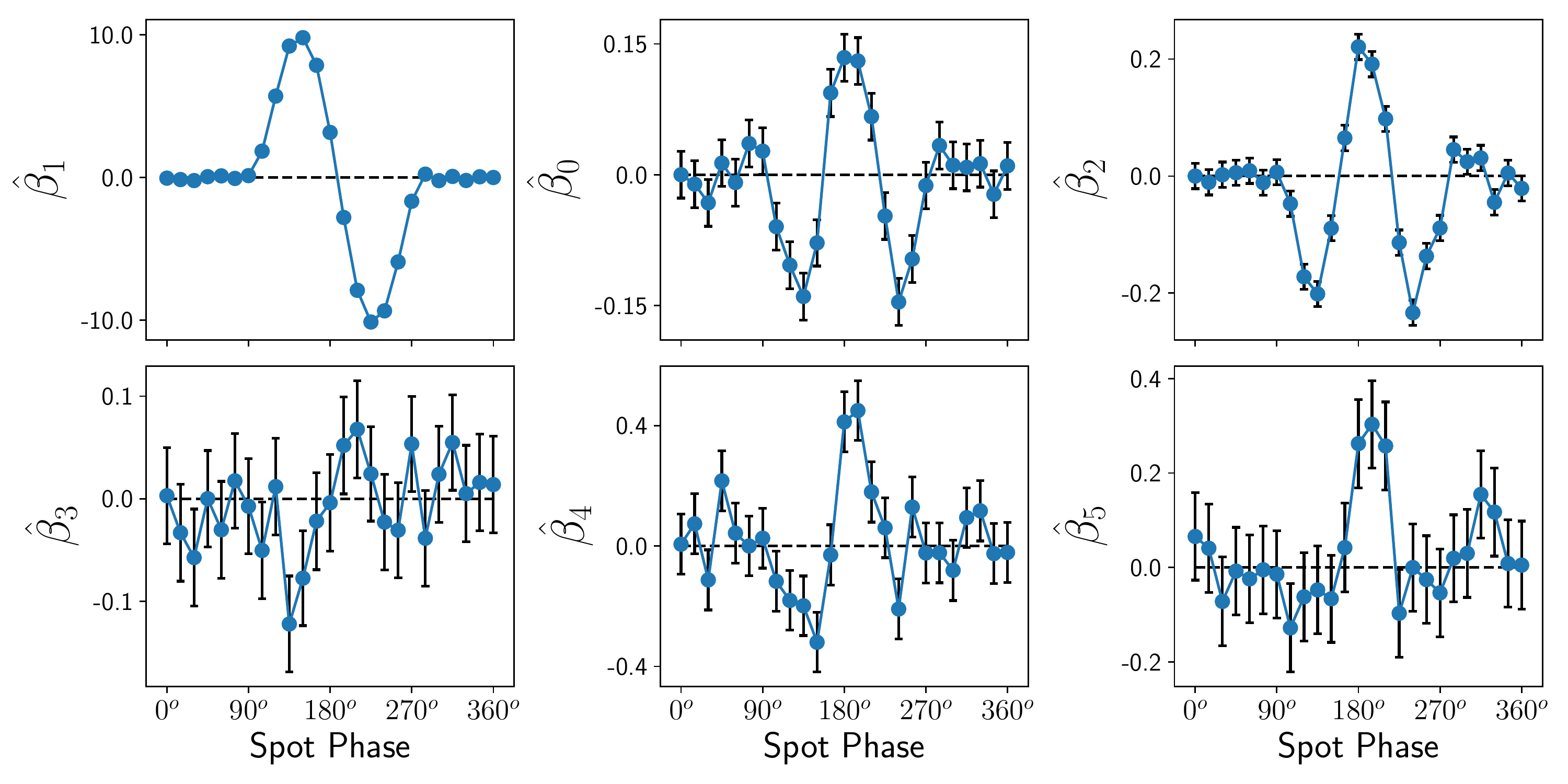}
\caption{The values of the fitted coefficients in Equation \eqref{eqn:full_mod2} for a $500$ SNR simulation of SOAP, plotted as a function of the rotational phase of the $1\%$ equatorial active region. Each panel corresponds to one of the coefficients, beginning in the top left with $\hat{\beta}_{1}$ which represents the estimated RV in \ms. All five other coefficients are those included in the F-test that produces the SAFE. Error bars are also shown for each coefficient and phase. Note that the error bars for $\hat{\beta}_{1}$ are plotted, but small enough to not be visible.}
\label{fig:coefs_vs_phase}
\end{figure}

Figures \ref{fig:exo_vs_spot} and \ref{fig:power_vs_snrphase} illustrate that when the active region is near phase $120^{\circ}$, $180^{\circ}$, or $240^{\circ}$, the power of the SAFE test is the highest. For further understanding of these phases, Figure \ref{fig:coefs_vs_phase} shows the estimated value of each coefficient in Equation \eqref{eqn:full_mod2} as a function of rotational phase according to a $500$ SNR simulation of SOAP.

As illustrated in Figure \ref{fig:coefs_vs_phase}, we find that the coefficients in Equation \eqref{eqn:full_mod2} with the largest t-statistic  when fitted to these simulations are $\beta_{0}$ and $\beta_{2}$, which reach local extrema at the three phases $120^{\circ}$, $180^{\circ}$, and $240^{\circ}$. This suggests that, after accounting for a Doppler shift, the most statistically significant characteristic of stellar activity according to SOAP is a symmetric change of absorption features.

We now compare the SAFE statistic to various other stellar activity indicator statistics in the literature, namely the BIS, BISlope, BiGauss, V-span, and FWHM. To do so we use the same spectra and simulation approach as we do for the SAFE above. Since these other indicators have not previously been used as hypothesis-test statistics for stellar activity in individual spectra, our first step is to estimate their null distribution. To do so we use $13000$ independent simulations of adding Poisson noise to the SOAP spectrum where no stellar activity is present.

\begin{figure}[ht!]
\centering
\includegraphics[scale=0.8]{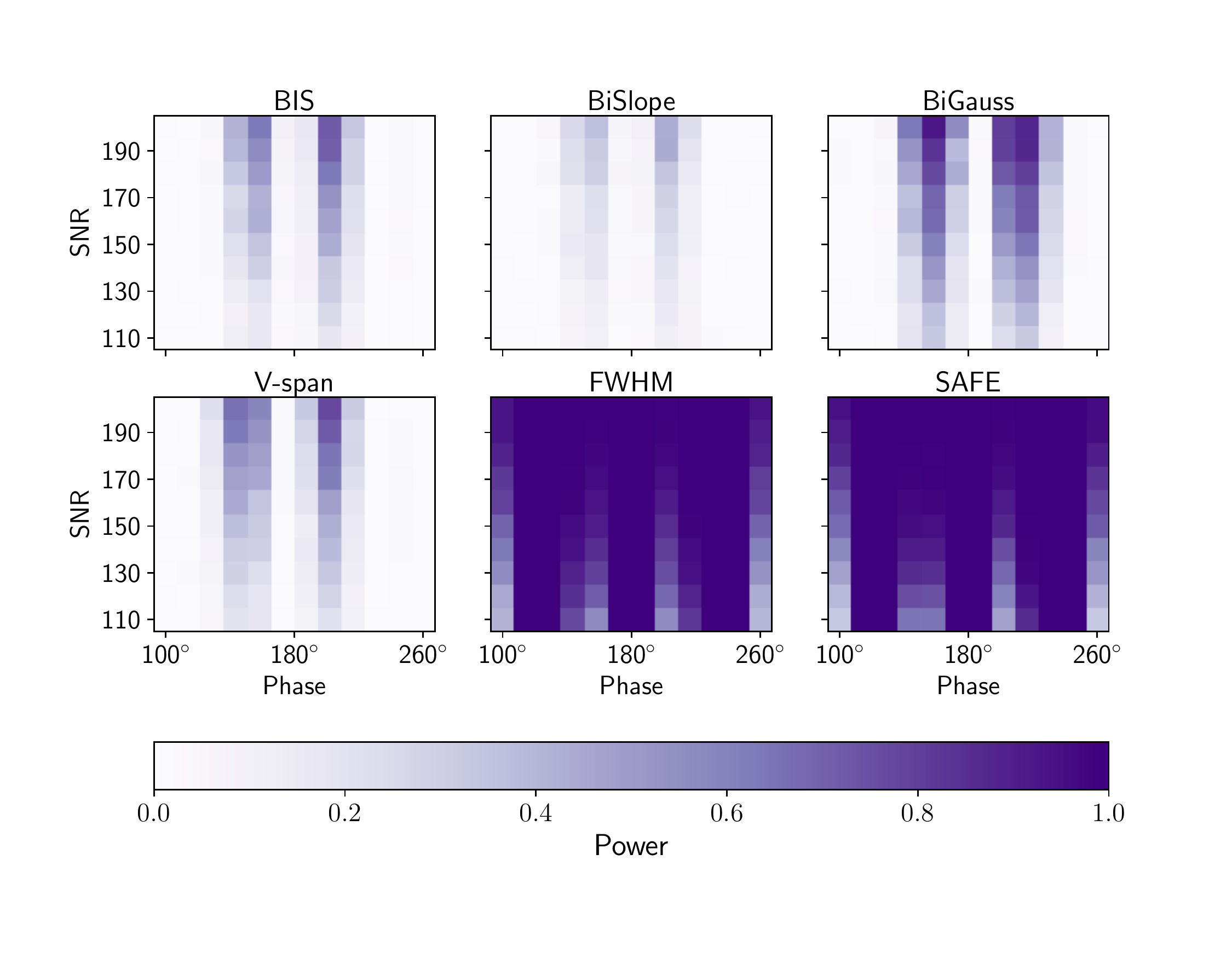}
\caption{Power comparison of the BIS, BISlope, BiGauss, V-span, and FWHM stellar activity indicators to the SAFE. For each indicator, we show the power as a function of the active region's phase and SNR. The phases where the active region is behind the star are not included.}
\label{fig:indicator_comp}
\end{figure}

For a given phase and SNR level, we estimate the power of using each indicator specified above to conclude that there is statistically significant stellar activity present. We do so with the proportion of $1000$ simulations that were either below the $0.5$ percentile, or above the $99.5$ percentile of the estimated null distribution (except in the case of the SAFE which is only statistically significant at large values, for which we calculate the proportion above the $99$ percentile. The power results are illustrated in Figure \ref{fig:indicator_comp}.

\begin{figure}[ht!]
\centering
\includegraphics[scale=0.7]{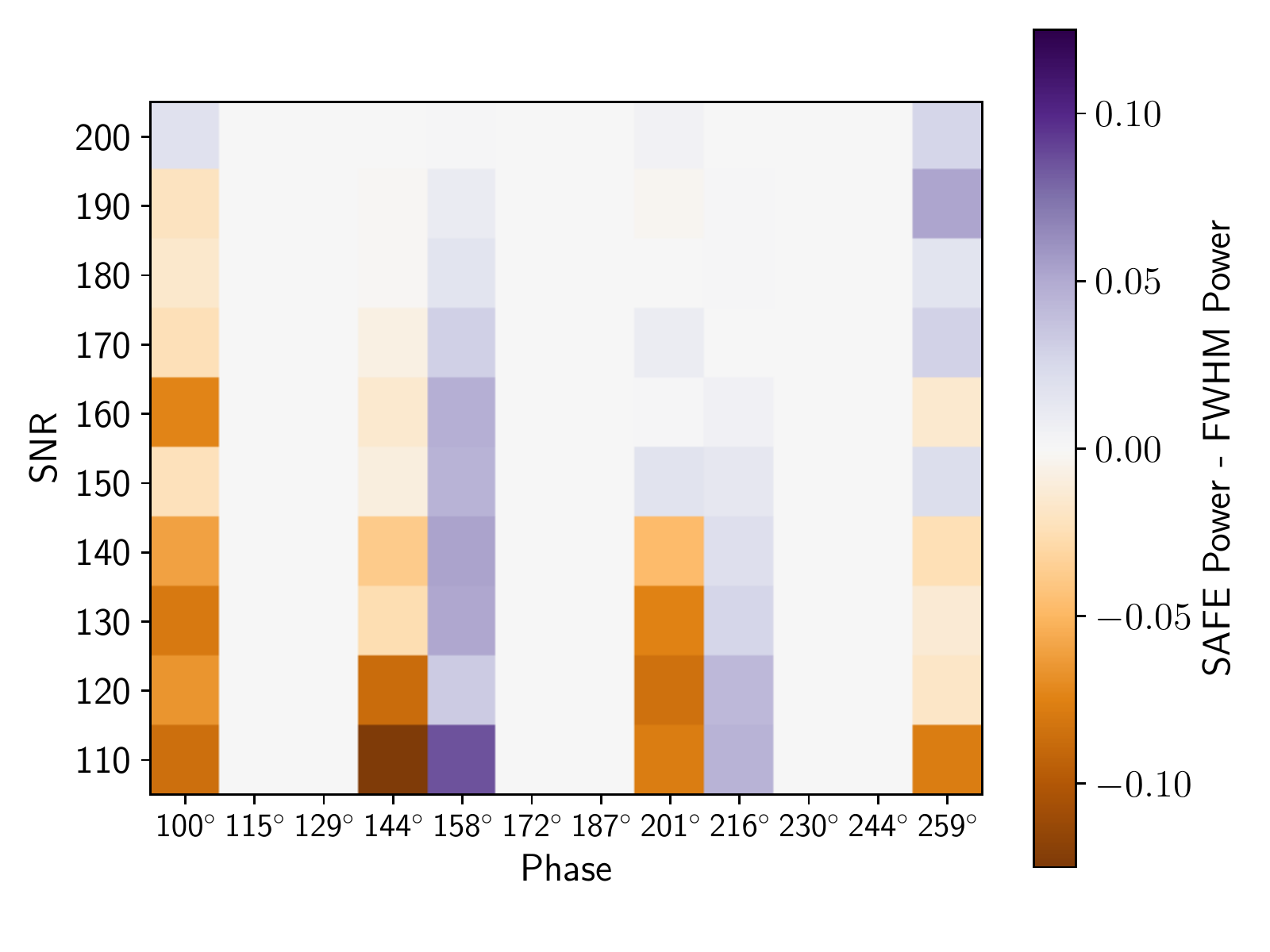}
\caption{Power comparison of the SAFE and the FWHM stellar activity indicators. The difference in statistical power between the SAFE and FWHM is indicated by the color bar on the right (that is centered at zero), and is shown as a function of the active region's phase and SNR.}
\label{fig:safe-fwhm_comp}
\end{figure}

Clearly, the FWHM and SAFE statistics have more power for detecting stellar activity than the BIS, BiSlope, BiGauss, and V-span in this simulation setting. This suggests that the most prominent effect of stellar activity, after accounting for an apparent Doppler shift, is a symmetric change to absorption features. Figure \ref{fig:safe-fwhm_comp} gives a power comparison of the SAFE to that of the FWHM by showing the power difference.

Figure \ref{fig:safe-fwhm_comp} illustrates that at various phases and low SNR the FWHM has higher power than the SAFE. But as the SNR increases to about $200$, the SAFE becomes more powerful for most phases.

\subsection{Template Estimation} \label{subsxn:template_est}

\begin{figure}[ht!]
\centering
\includegraphics[scale=0.7]{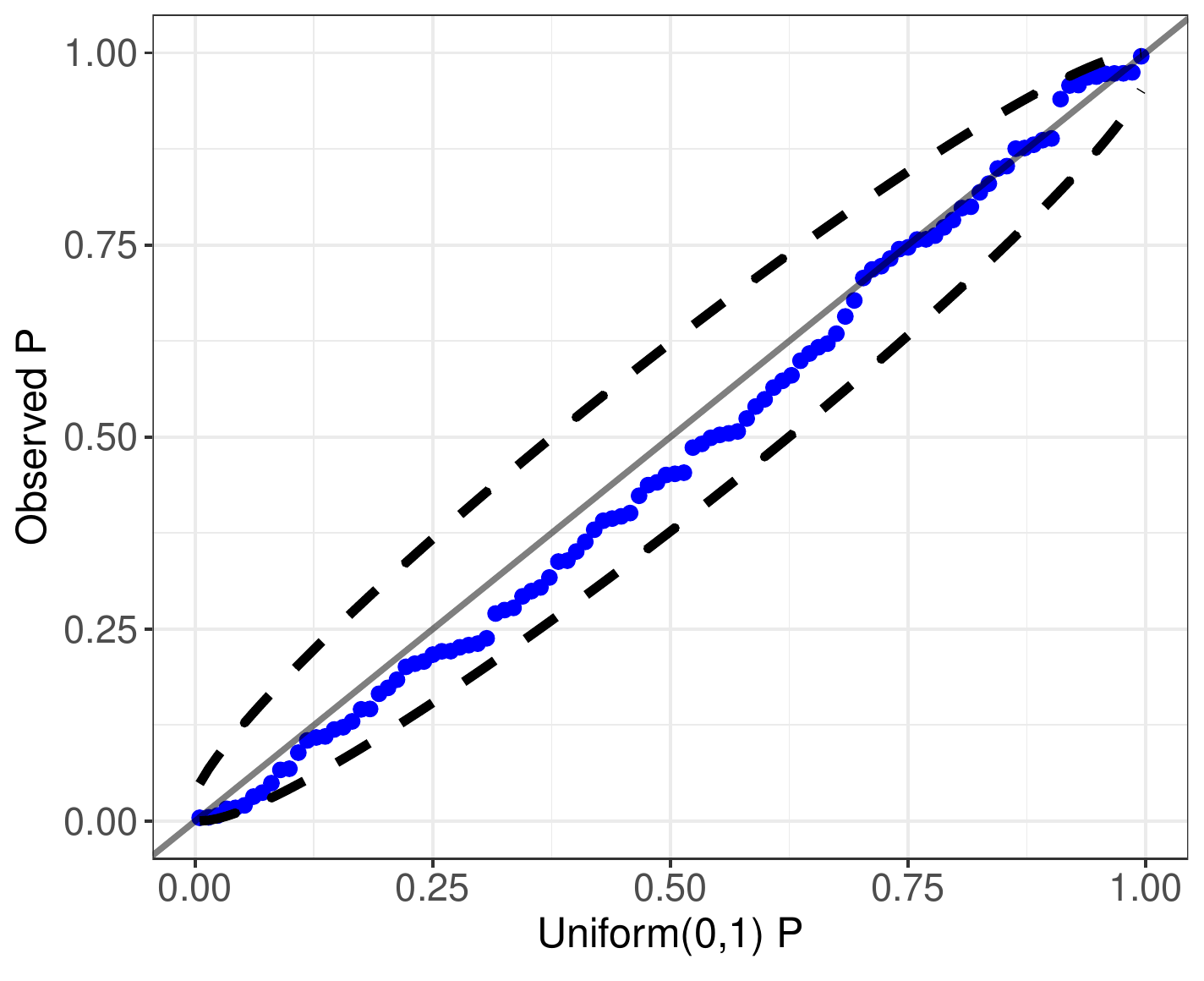}
\caption{Uniform quantile-quantile plot with the empirical p-value quantiles from the SAFE simulation on the vertical axis, and the corresponding Uniform(0,1) quantiles on the horizontal axis. $99$\% confidence bands are shown in dashed black.}
\label{fig:qqplot}
\end{figure}

We next simulate a set of spectra with the goal to check whether the SAFE null distribution is impacted by using an estimated template spectrum in place of the true template. To do so, we take the $106$ observed spectra of Tau Ceti (described below in Section \ref{sxn:applications}) and, for each time instance, replace the flux of each order with that of the corresponding wavelength region of a SOAP spectrum. The SOAP flux used for replacement has poisson noise with the same SNR as the corresponding Tau Ceti epoch. Depending on the epoch, these SNR range from $120$ to $250$, most of which are around $200$. It also has a Doppler shift representing the same RV as estimated by the CCF for the corresponding Tau Ceti epoch. After applying the full method to this simulation, the p-values shown on the vertical axis of the uniform quantile-quantile plot of Figure \ref{fig:qqplot} are given.\footnote{A uniform quantile-quantile plot shows the empirical quantiles of the sorted data on the vertical axis, plotted against the corresponding theoretical quantiles of the Uniform($0,1$) distribution. The closer the data lie to the diagonal in the plot, the more the data resemble a uniform distribution. For sorted p-values $p_{(1)} \leq p_{(2)} \leq ... \leq p_{(N)}$, the corresponding theoretical quantile of $p_{(i)}$ is $q_{i} = \left(i - 0.5\right)/N$. The values of the $100(1-\alpha)$\% confidence band at $q_{i}$ are given by the $\alpha/2$ and $1 - \alpha/2$ quantiles of the Beta($i, N-i+1$) distribution.} Since the coordinates do not step outside the $99$\% confidence bands, this gives support that the p-values for our simulation are approximately uniform. Therefore, we conclude that the template estimation does not significantly impact the null distribution of the SAFE. \changes{If the imperfect template estimation procedure had a strong influence on our ability to perform statistical inference about the SAFE, the data in Figure \ref{fig:qqplot} would depart significantly from the diagonal.}

\subsection{Realistic Stellar Activity} \label{subsxn:chris}
\begin{figure}[ht!]
\centering
\includegraphics[scale=0.78]{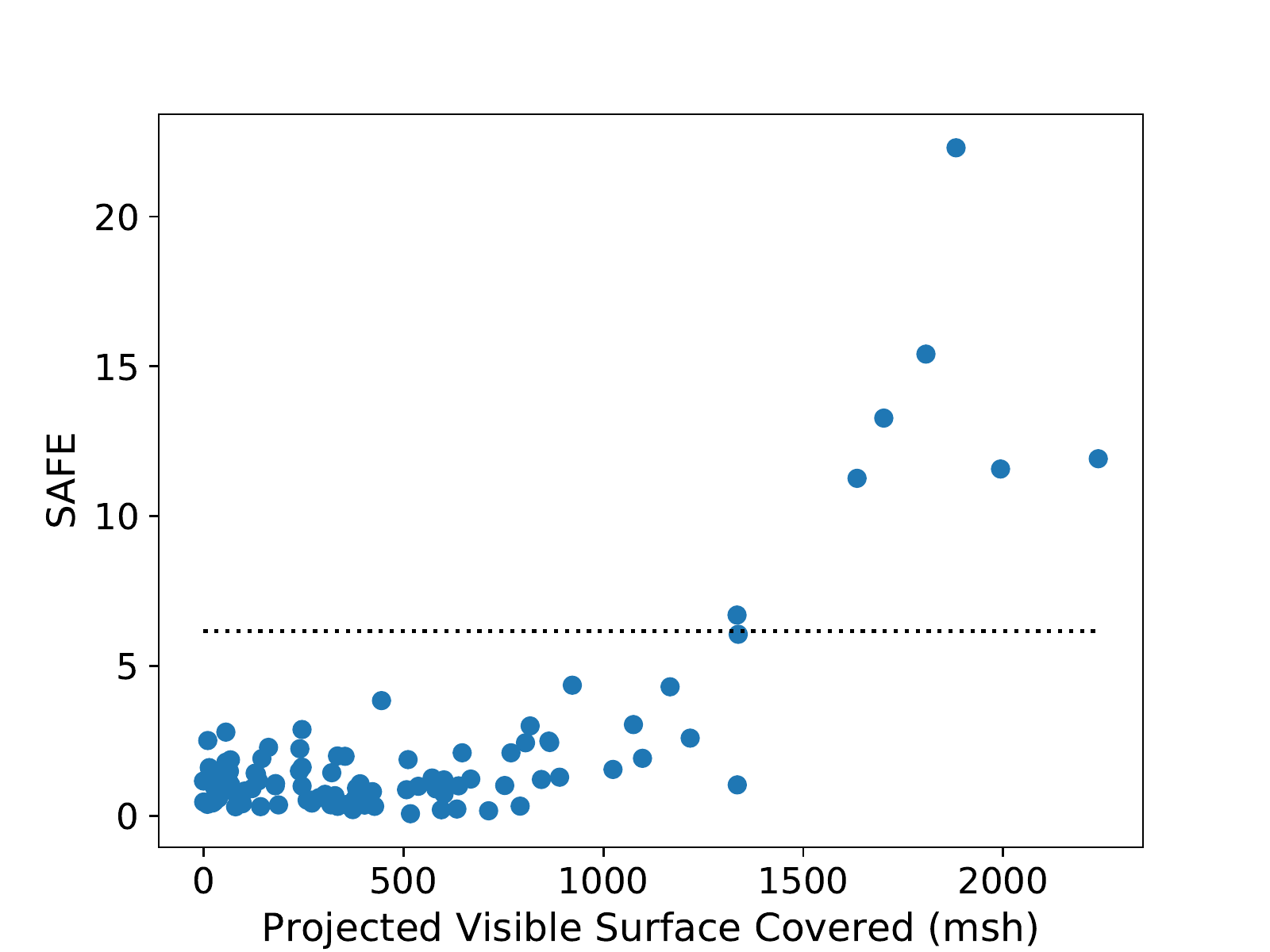}
\caption{The calculated value of the SAFE plotted against the true projected micro-solar hemispheres (msh) on the visible surface of the star covered by active regions. The horizontal dashed line represents the critical value of SAFE after correcting for multiple testing, above which the p-value is less than $0.001$.}
\label{fig:safe-chris}
\end{figure}

Activity on stars often include multiple active regions with variable sizes that change dynamically.  We apply the proposed method to simulated spectra from \citet{gilbertson20} that were generated using a more realistic active region model. The apparent RV induced by these models is on the scale of about $1$ \ms, characteristic of stellar variability of the Sun. We randomly sample $100$ of the $730$ available epochs, reduce the resolution of each to $137500$ to match the reported resolution of EXPRES \citep{petersburg2020}, limit to the wavelength region between $5000$ and $6500$ \AA, and decrease the SNR to $250$. Using these $100$ spectra we estimate the template spectrum and calculate the SAFE after subtracting off the fit of the median estimated coefficients over time. 
Figure \ref{fig:safe-chris} compares the SAFE to the area in micro-solar hemispheres (msh) of the visible stellar surface covered by active regions, projected onto the visible disk.

Figure \ref{fig:safe-chris} demonstrates that, under more realistic active region effects on Sun-like stars, the SAFE becomes statistically significant when the percent of the visible projected disk of the star covered by active regions is greater than approximately $0.15\%$. It also shows a noticeable monotonic association between the SAFE and the projected visible surface covered.

\section{Applications to EXPRES Data} \label{sxn:applications}

To test the performance of the SAFE on real data we use spectra recently collected by EXPRES for the stars HD 10700 (Tau Ceti), HD 22049 (Eps Eri), and HD 3651. Eps Eri is known to be a star with relatively high stellar variability \citep{kelch78, simon80, thatcher91, gray95}, whereas Tau Ceti \citep{gray84, saar97, frick97} and HD 3651 \citep{frick97} are known to be relatively quiet. Our dataset consists of $106$ spectra for Tau Ceti, $60$ spectra for Eps Eri, and $52$ spectra for HD 3651. These spectra were all collected in the later half of 2019.

Each of these spectra come with a barcentric corrected wavelength solution, normalized flux, and telluric models. Uncertainties for the normalized flux are also provided, which we use to determine the weights in the linear model for the SAFE.

\subsection{Data Preprocessing}

To mitigate the effects of tellurics, we mask out all pixels of each spectra where the telluric model normalized flux generated from selenite \citep{leet19} is less than $1.0$. 
We also use cubic spline interpolation to stitch the orders together, giving a single two-dimensional array of normalized flux and wavelength for each spectrum. We further limit the stitched spectra to be between $5000$ and $7000$ \AA\ as the wavelength solution is known to be increasingly less accurate outside this range.

Although the spectra are provided with the continuum normalized, we find that an additional normalization step is needed to remove remaining trends in the continuum, particularly near the edges of orders. To further adjust the normalization the Absorption Feature Finder (AFF) algorithm of \citet{holzer20} is used with a small width parameter to identify as many potential absorption features as possible. The portions of the spectra that do not lie inside any of the detected absorption feature wavelength bounds are then used as windows of the continuum. Because the AFF algorithm can sometimes miss absorption features, any of the remaining pixels that have a normalized flux more than one multiple of the interquartile range below the $25$th percentile of the remaining subset are removed. A running median is then applied to estimate the remaining continuum to be divided out. 
Finally, we adjust the wavelength axis of each spectrum by dividing out the apparent Doppler shift using the RV estimated by the EXPRES pipeline \changes{which, as previously noted, need only be within $5$ \ms\ of the true velocity}. This allows for a more accurate estimation of the template spectrum (described in Section \ref{subsxn:template_est}) that has minimal broadening of absorption features.

\subsection{Calculating SAFE}

For each of the spectra we first estimate a template spectrum using the nonparametric approach described in \citet{holzer20}. Because replacing hardware or other components of the instrumentation can effect the spectrum in ways that lead to a statistically significant value of SAFE, separate estimates of the template spectrum before and after such changes is recommended. Such changes took place in August 2019 following $47$ (of the total $106$) observations of Tau Ceti. These updates helped improve the stability of the laser frequency comb by replacing the photonic crystal fiber and changing the polarization of the light. (All observations of Eps Eri and HD 3651 came after these changes.) Therefore, we estimated two separate template spectra for Tau Ceti using the approach described in Section \ref{subsxn:template_est}.

For each of these template spectra, we obtain estimates $\hat{\mu}_{i}$ and $\hat{\sigma}_{i}$ for each feature using the process described at the beginning of Section \ref{subsxn:single_feature}.\footnote{For the HD 10700 template spectrum before the instrumental changes in August 2019, $1191$ of the $2297$ features remained after these criteria. For the HD 10700 template spectrum after the instrumental changes, $1128$ of the $2172$ features remained. Out of the $2302$ features for HD 3651, $871$ remained. And HD 22049 had $877$ of its $2309$ features remain.} 

We then perform weighted least squares with the difference flux according to Equation \eqref{eqn:full_mod2}. Occasionally, we find that there are strong outliers present in the regression, mostly due to absorption features with a poor algorithmic convergence of fitting a Gaussian and failing to be eliminated as described in the preceding paragraph. Consequently, we also remove points at time $t$ that have an influence\footnote{For a linear model $Y = X\beta + \varepsilon$, with design matrix $X$ and response vector $Y$, the vector of fitted values is $\hat{Y} = X \left(X^{T}X\right)^{-1}X^{T}Y = HY$. The influence of the $i$'th data point, which will be relatively far from zero if $X_{i,\cdot}$ is an outlier, is given by the expression $\dfrac{\left( Y_{i} - \hat{Y}_{i} \right) H_{i,i}}{1 - H_{i,i}}$. } greater than $50$ multiples of the interquartile range above the $75$'th percentile of all the influence values at time $t$. We find that such strong influence points rarely occur and are usually due to cosmic rays. Before calculating the SAFE according to Equation \eqref{eqn:SAFE_def} for all observed spectra we make one final adjustment to the template spectrum by taking the median value of each fitted coefficient $\hat{\beta}_{i,t}$ across time, calculating the median difference flux from the median coefficients, and subtracting the median difference flux from the estimated template. This helps to further correct for the bias present in the estimated template spectrum. Using this adjusted template to calculate the difference flux for each normalized spectrum, we then fit the coefficients and calculate the $\mathrm{SAFE}_{t}$ at each time $t$.

\subsection{Results}

\begin{figure}
\centering
\includegraphics[scale=0.85]{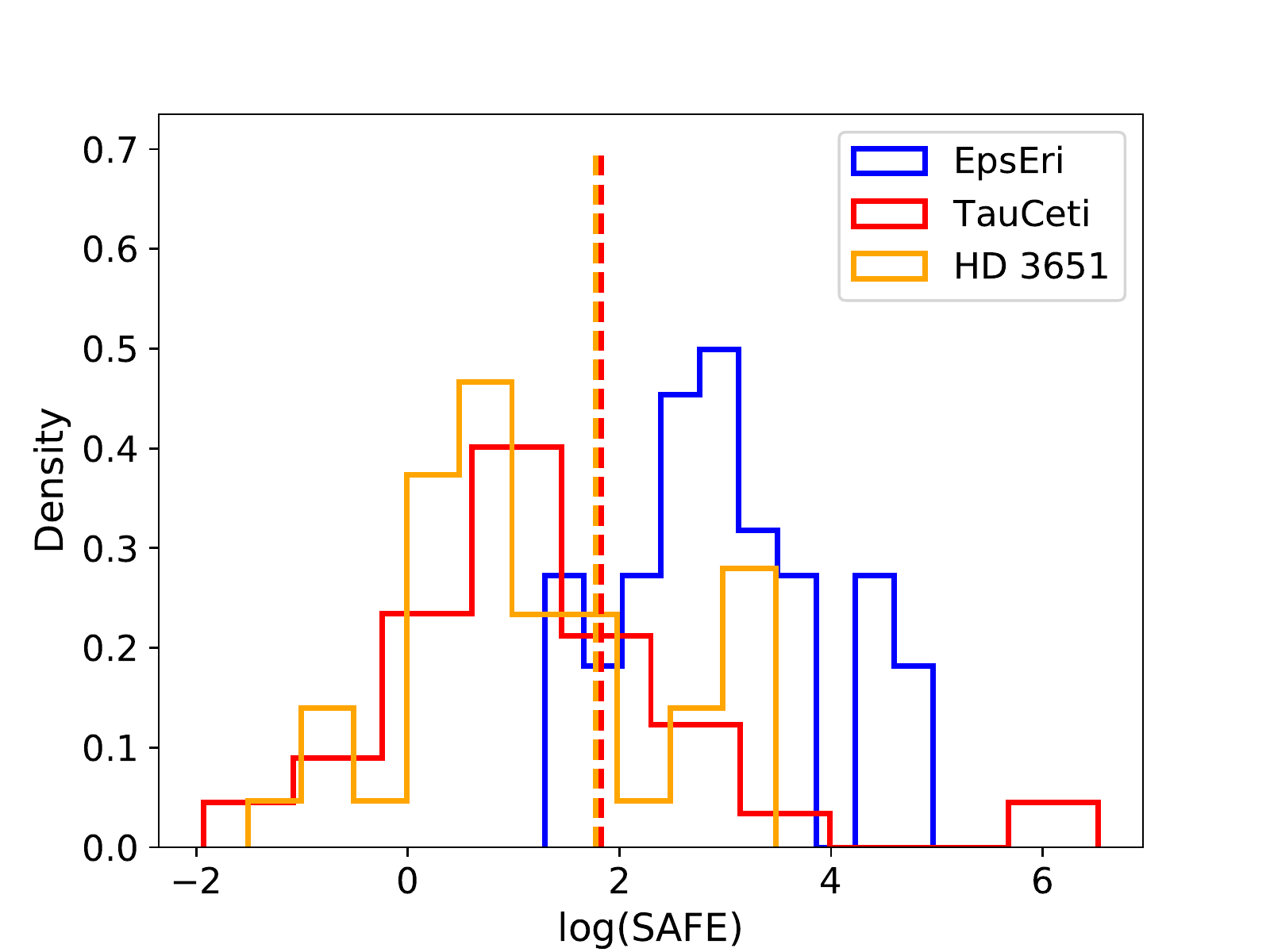}
\caption{Histograms representing the distribution of the natural log of SAFE for the three stars Eps Eri (blue), Tau Ceti (red), and HD 3651 (yellow). The (multiple testing corrected) critical value for a significance level of $0.001$, above which represents statistically significant values of SAFE, is shown by the vertical dashed line. Note that all three stars appear to have approximately the same critical value in this plot. }
\label{fig:hd3651hd10700hd22049comp}
\end{figure}

Figure \ref{fig:hd3651hd10700hd22049comp} shows the distribution of the (log) SAFE for Eps Eri, Tau Ceti, and HD 3651. This illustrates that, as expected, Tau Ceti and HD 3651 mostly have values of SAFE that are not statistically significant, occasionally showing a significant stellar activity signal. On the other hand, the large majority of Eps Eri observations gave a value of SAFE above the critical value.

\begin{figure}[ht!]
\centering
\includegraphics[scale=0.85]{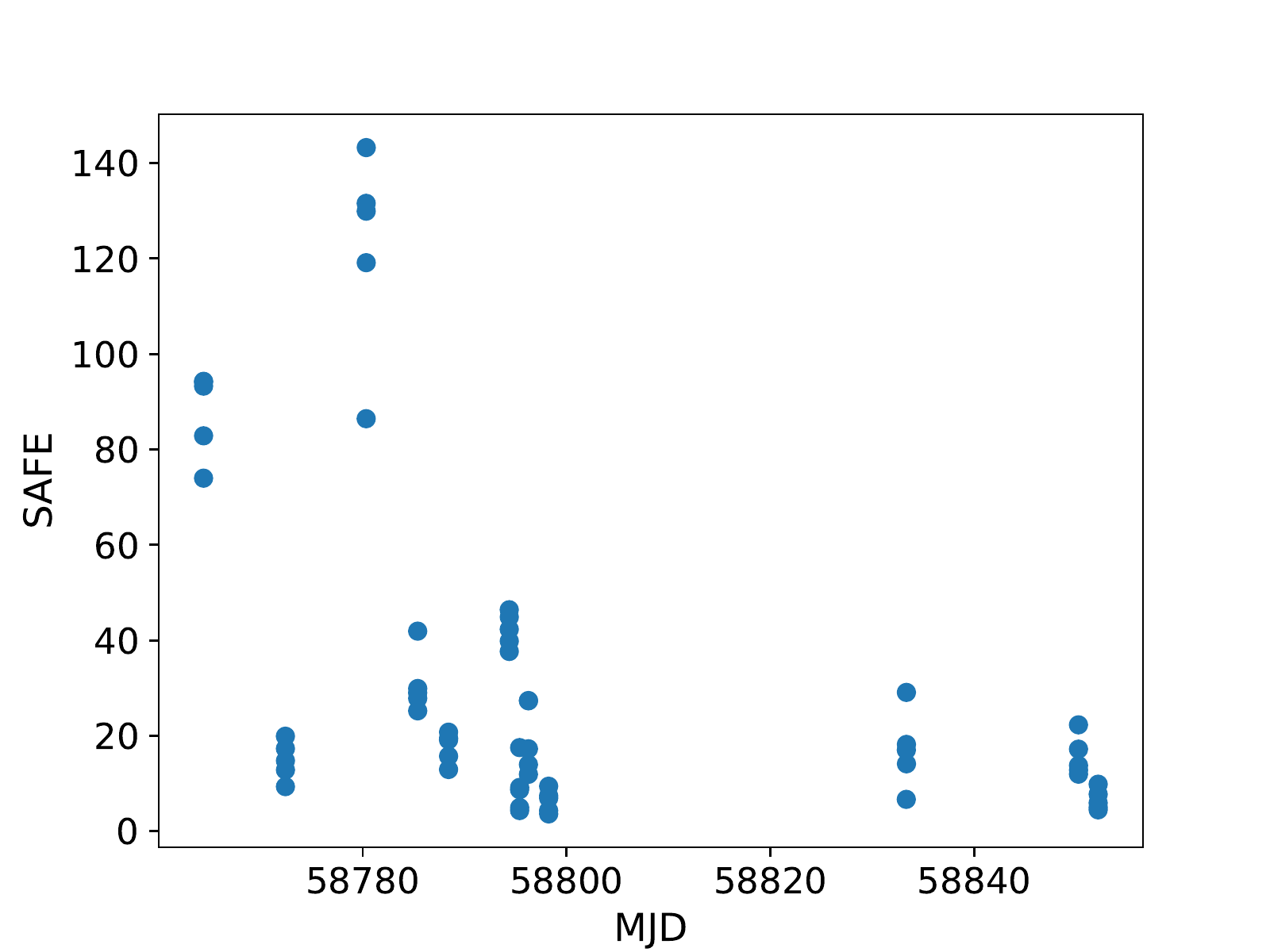}
\caption{Scatterplot of the SAFE against the (barycentric) MJD for observations of Eps Eri (HD 22049). The scatter between nights is greater than the average scatter within nights, which is expected for stellar activity indicators.}
\label{fig:safe_vs_mjd}
\end{figure}

As further evidence that the SAFE statistic is detecting stellar activity in the spectra for Eps Eri, Figure \ref{fig:safe_vs_mjd} shows the SAFE plotted against the Modified Julian Date (MJD). Since stellar variability in the form of active regions changes on the order of days, we would expect to find similar values of SAFE for observations taken in the same night.

For Eps Eri we find that the standard deviation of the SAFE between nights ($34.5$) is approximately six times larger than the standard deviation within nights ($5.78$). Furthermore, we did not find any noticeable association between the SAFE statistic and any of the weather conditions provided by EXPRES: cryostat temperature, cryostat pressure, angular distance of the moon, angular distance of the Sun, airmass, and exposure time.

\changes{
We also analyzed the star HD 34411, one of the most frequently observed stars in the EXPRES pipeline. With a total of $173$ spectra ranging from (barycentric) MJD $58437$ to $59135$ we calculated the SAFE with the same procedure previously described with initial RV estimates given by the method of \citet{holzer20}. A total of $24$ of the observations produced a statistically-significant value of the SAFE.

As a first step of extending the use of the SAFE for correcting the RV measurements to be less influenced by stellar activity, we suggest simply removing observations that we conclude have a statistically-significant stellar activity signal present.\footnote{We acknowledge that this approach of simply eliminating observations is likely to not work well on very active stars where nearly all observations have a statistically-significant stellar activity signal.} For HD 34411, by eliminating the $33$ observations where the SAFE is above the critical value or occur on the same night as an observation with a very large value of SAFE (i.e., greater than $10$), we find that the root mean squared deviation from the average RV decreases from $2.08$ \ms\ to $1.72$ \ms. This represents an improvement of estimated RV precision of approximately $17\%$. 
}

\section{Discussion}

The proposed SAFE statistic aims to detect any shape change in spectral absorption features that is not due to a Doppler shift. Under the assumption that the line spread function is stable and all instrumental effects have been properly corrected for, the only remaining source of absorption feature shape changes is stellar variability in the observed star. In this sense, the SAFE can be thought of as a combination of many classical stellar activity indicators. This is because it is composed of a linear combination of Hermite-Gaussian functions, $\psi_{n}$. For a single absorption feature a change in the line depth would be characterized with $\psi_{0}$. $\psi_{2}$ would represent a change in the line width, similar to the FWHM. A change in skewness would result in some linear combination of $\psi_{3}$ and $\psi_{5}$, similar to the BIS, BiSlope, V-span, and BiGauss. And $\psi_{4}$ helps to represent a change in line kurtosis. Therefore, instead of trying to detect a specific type of change due to stellar activity, the SAFE aims to detect any change in line shape that is not due to a simple Doppler rescaling of the wavelength axis.

Compared to the many other stellar activity indicators, the SAFE is unique in that it's null distribution is well known. Consequently, it allows for testing the hypothesis that no stellar activity is present. 

One of the benefits of building the SAFE with a linear model comes from the fact that rarely, if ever, is there just one active region or form of stellar activity on a given star. Under the assumption that the cumulative effect of multiple active regions is just the superposition of each individual active region, a linear model allows for not knowing the number of active regions in advance. Active regions are also known to vary in size. Assuming that the effect of a different sized active region would just be a rescaling of the difference flux imposed on each absorption feature, a linear model also allows for not knowing the size of each active region. 

Unlike in \citet{holzer20}, we do not assume absorption features are Gaussian. \changes{This allows for using spectral lines with significantly broadened wings. It also means we do not need to avoid blended spectral lines.} Instead we assume that each absorption feature's difference flux due to a Doppler shift can be decomposed as a linear combination of the Hermite-Gaussian functions up to degree $5$. Consequently, we are unable to mathematically quantify how small of an RV is small enough for our assumptions to be met; our simulations suggest that a RV of $10$ \ms\ is within the acceptable range. Furthermore, the Hermite-Gaussian functions are not the only basis one could choose for calculating the SAFE. We prefer this basis because it is smooth, localized, and composed of even and odd functions.

As mentioned above, the SAFE is calculated from using a large portion of the spectrum. By dropping the assumption of Gaussian-shaped absorption features, we are able to use more features found by the AFF algorithm compared to when the Gaussian-shape assumption is used. While we do use the continuum in our renormalization as part of the data preprocessing, only pixels corresponding to absorption features are used in the regression model used for calculating the SAFE.

The dimensional reduction we perform using rPCA over each degree of Hermite-Gaussian coefficients for SOAP spectra can be considered a way of quantifying how strongly certain absorption features are affected compared to others. The ability of the SAFE to detect statistically-significant stellar activity signals in Eps Eri spectra indicate that the relative amplitudes given by rPCA on SOAP spectra are accurate enough to extend to real data. However, it remains unknown how well such relative amplitudes will do at detecting stellar variability in stars that are very different than the Sun. Furthermore, the amplitudes may need to be adjusted if one aims to detect other forms of stellar variability than active regions. For example, since very few of the observations of Tau Ceti gave a statistically significant SAFE, but all were (intentionally) collected in a short time interval that likely did not wash out the signal of pressure mode oscillations \citep{chaplin19}, this suggests that the relative amplitudes are not tuned to detecting this type of stellar variability.

\begin{figure}[ht!]
\centering
\includegraphics[scale=0.78]{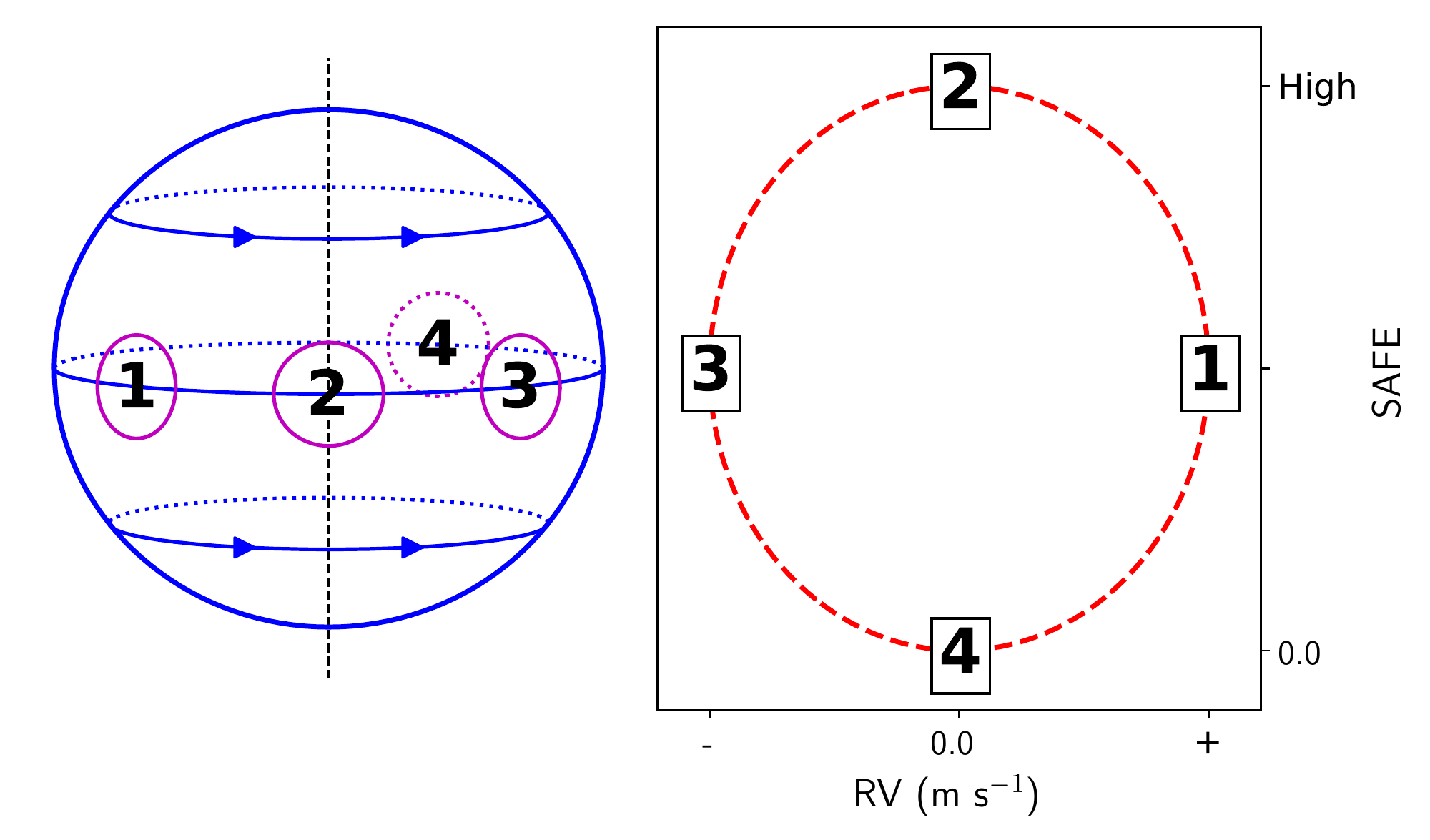}
\caption{Visual graphic representing qualitatively why we do not expect the SAFE to be correlated with the apparent RV due to an active region. The sphere on the left represents a star with four sequential phases of an equatorial active region. Solid lines represent the visible side of the star and dotted lines represent the back. The plot on the right shows the expected behavior of the SAFE and RV for each of the four phases. For a different sized active region, the coordinates of the four points in the right panel would simply be scaled by a certain factor. Note that state \textbf{4} is for when the active region is completely on the non-visible side of the star.
}
\label{fig:rv_vs_safe}
\end{figure}

For the simulation studies of SAFE, as well as the applications to stars observed by EXPRES, we find that the coefficient of $\Psi_{2}$ is the most statistically significant when calculating the SAFE. This suggests that the most noticeable difference between a Doppler shift and a signal from an active region is the symmetric change represented by a multiple of $\psi_{2}$. Therefore, as illustrated in Figure \ref{fig:rv_vs_safe}, we do not expect the SAFE to be correlated with the apparent RV due to an active region. 

Because the SAFE is not expected to be correlated with the apparent RV, it is primarily designed to detect, rather than directly correct for, the presence of stellar activity. However, the SAFE or the individual fitted coefficient values of Equation \eqref{eqn:full_mod2} may still be used to help correct the apparent RV by including it in a time-series model such as that of \citet{rajpaul15} or \citet{jones17}.

\changes{While the approach for cleansing the RV measurements of stellar activity by simply removing observations with a large value of SAFE may be useful for stars with intermittent activity, it would result in rejecting a large fraction of data for other stars that frequently have detectable levels of stellar activity.  Therefore, future studies should investigate whether the SAFE statistic or the $\hat{\beta}_{d}$ coefficients could be used to predict the RV perturbation due to stellar activity and provide an improved estimate for the RV.  As a first step in this direction, we investigated the ability of these coefficients calculated from half a year of simulated solar data \citep{gilbertson20} to perform this task. It is unclear how the coefficients will map onto RV contamination, so we fit a version of the multivariate GP model described in \citet{gilbertson20toward} to the $\hat{\beta}_{1}$ coefficients (which represent initial RV estimates) and the first two series of PCA scores created from the $\hat{\beta}_{0}$, $\hat{\beta}_{2}$, and $\hat{\beta}_{3}$ components. We found that, once the model hyperparameters were fit, conditioning the posterior of the model on only the stellar activity information (i.e., no RV information) removed over half of the variance in the RV measurements contributed by the stellar variability. Thus, the use of SAFE coefficients to improve the precision of RV measurements in the presence of stellar variability appears to be a promising avenue for future research.}

A number of improvements could potentially be made to the methodology of calculating the SAFE. The assumption made about the noise being independent is not necessarily true due to the point spread function of the spectrograph. Adjusting for this by using a non-diagonal covariance matrix in the linear model of Equation \eqref{eqn:full_mod2} would help account for this. Another improvement would be to include $\Psi_{1}$ in the model and test if the coefficient of this additional term is also simultaneously zero. This would potentially allow the SAFE to more easily detect antisymmetric shape changes in absorption features due to stellar variability. While $\Psi_{1}$ is not included due to potential collinearity, this issue may be avoidable by formulating the regression in a Bayesian framework. Including $\Psi_{d}$ for $d > 5$ may also help improve the SAFE.

While a good number of absorption features were removed in this work, the SAFE has the potential to be calculated from nearly all spectral features. If the continuum normalization were to be done with high accuracy, and complicated absorption features could be well modeled, then many more such features could be included in the linear model. This could increase the statistical power of the SAFE even further.

Finally, an improvement that we suspect will greatly improve upon this work is to account for effects of stellar activity that are not shared across most absorption features. Our use of relative amplitudes in building the explanatory variables for the regression model of Equation \eqref{eqn:full_mod2} essentially limits the SAFE to detecting spectrum-wide effects. But it is well known that some absorption features are much more sensitive to stellar activity than others \citep{dumusque18,wise18,ning19}. Including additional terms in Equation \eqref{eqn:full_mod2} for such activity-sensitive spectral lines could allow the SAFE to detect both spectrum-wide and feature-specific effects simultaneously.

\section{Conclusion}

We introduce a new stellar activity indicator called the SAFE, which is an F-statistic for a linear model of a spectrum's difference flux on sums of Hermite-Gaussian functions centered on absorption features. Through simulation we demonstrate that the SAFE follows the assumed F-distribution, is not affected by a Doppler shift, and is associated with the area of the visible projected surface covered by active regions. We also compare the statistical power of many classical stellar activity indicators and find that the SAFE and FWHM perform the best. Furthermore, the SAFE is estimated on a set of spectra recently collected by EXPRES from HD 22049, HD 10700, and HD 3651. HD 22049 is known to be very active and results in statistically significant values of SAFE in almost all of the spectra. Additionally, the other two stars, known to be relatively inactive, only occasionally have statistically significant stellar activity detected by the SAFE. Overall, the SAFE is a new stellar activity indicator that detects stellar variability with high power and \changes{appears to show promise of being} useful in disentangling such signals from exoplanet-induced Doppler shifts. Data and code associated with this work is accessible in the following repository: \href{https://github.com/parkerholzer/safe_statistic}{https://github.com/parkerholzer/safe\textunderscore
statistic}. \changes{The methodology is also implemented in the R package \textbf{rvmethod 0.2.0} available on the CRAN.}

\acknowledgments
This work used data from the EXtreme PREcision Spectrograph (EXPRES) that was designed and commissioned at Yale with financial support by the U.S. National Science Foundation under MRI-1429365 and ATI-1509436 (PI D. Fischer).  We gratefully acknowledge support for telescope time using EXPRES at the LDT from the Heising Simons Foundation and an anonymous Yale donor. DAF, JCK, EBF and CG acknowledge critical support for investigation of photospheric noise in RV data from the NSF AST-1616086 and NASA 80NSSC18K0443. LLZ gratefully acknowledges support from the NSF GRFP. We thank the Yale Center for Research Computing for guidance and use of the research computing infrastructure.
E.B.F. acknowledges the Penn State Center for Exoplanets and Habitable Worlds, which is supported by the Pennsylvania State University and the Eberly College of Science.
Computations for this research were performed on the Pennsylvania State University’s Institute for Computational and Data Sciences’ Roar supercomputer).
This research was supported by Heising-Simons Foundation Grant \#2019-1177 (E.B.F.). 
This work was supported by a grant from the Simons Foundation/SFARI (675601, E.B.F.).
E.B.F. acknowledges the support of the Ambrose Monell Foundation and the Institute for Advanced Study.

\bibliography{HGRVNotes}
\bibliographystyle{aasjournal}


\end{document}